\documentclass[twoside,english,pra,showkeys,showpacs,reprint,notitlepage,onecolumn,superscriptaddress,nofootinbib]{revtex4-1}
\usepackage[T1]{fontenc}
\usepackage[latin9]{inputenc}
\usepackage{geometry}
\usepackage{array}
\usepackage{booktabs}
\usepackage{units}
\usepackage{multirow}
\usepackage{amsbsy}
\usepackage{amstext}
\usepackage{graphicx}

\makeatletter

\providecommand{\tabularnewline}{\\}

\usepackage{color}
\definecolor{blueviolet}{rgb}{0.2, 0.2, 0.6}
\usepackage[pdftex,          
    bookmarks=false,          
    colorlinks=true,
    allcolors=blueviolet,
    pdfstartview={FitH},  
    ]{hyperref}

\usepackage{bm}
\usepackage{MnSymbol,amsfonts,mathtools}
\usepackage{graphicx} 
\usepackage{tikz} 
\usepackage{cases}     

\makeatother

\usepackage{babel}
\begin{document}
\global\long\def\bra{\langle}
\global\long\def\ket{\rangle}
\global\long\def\half{\frac{1}{2}}
\global\long\def\c{\pmb{s}}
\global\long\def\r{\pmb{m}}
\global\long\def\dfs{\textnormal{\textsc{dv}}}
\global\long\def\ft{\textnormal{\textsc{cv}}}
\global\long\def\fs{\textnormal{\textsc{rot}}}
\global\long\def\pr{\prime}
\global\long\def\x{\pmb{x}}
\global\long\def\p{\pmb{p}}
\global\long\def\lket#1{\left|#1\right\rangle }
\global\long\def\an{\Phi}
\global\long\def\ph{\pmb{n}}
\global\long\def\NN{\pmb{N}}
\global\long\def\k{\pmb{\theta}}
\global\long\def\o{\omega}
\global\long\def\aa{\pmb{a}}
\global\long\def\dg{\dagger}
\global\long\def\O{\Omega}
\global\long\def\Z{\mathbb{Z}}
\global\long\def\bb{\textnormal{\ensuremath{\mathbf{b}}}}
\global\long\def\X{\pmb{X}}
\global\long\def\P{\pmb{q}}
\global\long\def\yy{\pmb{y}}
\global\long\def\ZZ{\pmb{Z}}
\global\long\def\Y{\pmb{Y}}
\global\long\def\s{\sigma}
\global\long\def\V{\pmb{V}}
\global\long\def\U{\pmb{U}}
\global\long\def\R{\pmb{F}_{\dfs}^{2}}
\global\long\def\A{\hat{A}}
\global\long\def\B{\hat{B}}
\global\long\def\T{\hat{T}}
\global\long\def\C{\hat{C}}
\global\long\def\d{\delta}
\global\long\def\a{\alpha}
\global\long\def\b{\beta}
\global\long\def\N{{\cal N}}
\global\long\def\con{{\cal C}}
\global\long\def\F{\pmb{F}}
\global\long\def\D{\pmb{D}}

\global\long\def\ci{s}
\global\long\def\ri{m}
\global\long\def\M{\mathcal{M}}
\global\long\def\L{\mathcal{L}}
\global\long\def\tr{\text{Tr}}
\global\long\def\id{\boldsymbol{1}}
\global\long\def\Re{\text{Re}}
\global\long\def\Im{\text{Im}}

\DeclareRobustCommand{\upl}{\resizebox{10pt}{!}{\!\!\!\!
\begin{tikzpicture}[line join=round,x=5pt,y=5pt]   
\draw[black] (-0.5,-0.5)--(-0.5,0.5)--(0.5,0.5)--(0.5,-0.5)--(-0.5,-0.5);
\draw(0,0.5)        node{\textbullet};
\draw(0.5,0)        node{\phantom{\textbullet}};
\draw(0,-0.5)        node{\phantom{\textbullet}};
\draw(-0.5,0)        node{\phantom{\textbullet}};
\end{tikzpicture}}}
\DeclareRobustCommand{\rpl}{\resizebox{10pt}{!}{\!\!\!\!
\begin{tikzpicture}[line join=round,x=5pt,y=5pt]   
\draw[black] (-0.5,-0.5)--(-0.5,0.5)--(0.5,0.5)--(0.5,-0.5)--(-0.5,-0.5);
\draw(0,0.5)        node{\phantom{\textbullet}};
\draw(0.5,0)        node{\textbullet};
\draw(0,-0.5)        node{\phantom{\textbullet}};
\draw(-0.5,0)        node{\phantom{\textbullet}};
\end{tikzpicture}}}
\DeclareRobustCommand{\lpl}{\resizebox{10pt}{!}{\!\!\!\!
\begin{tikzpicture}[line join=round,x=5pt,y=5pt]   
\draw[black] (-0.5,-0.5)--(-0.5,0.5)--(0.5,0.5)--(0.5,-0.5)--(-0.5,-0.5);
\draw(0,0.5)       node{\phantom{\textbullet}};
\draw(0.5,0)        node{\phantom{\textbullet}};
\draw(0,-0.5)      node{\phantom{\textbullet}};
\draw(-0.5,0)        node{\textbullet};
\end{tikzpicture}}}
\DeclareRobustCommand{\dpl}{\resizebox{10pt}{!}{\!\!\!\!
\begin{tikzpicture}[line join=round,x=5pt,y=5pt]   
\draw[black] (-0.5,-0.5)--(-0.5,0.5)--(0.5,0.5)--(0.5,-0.5)--(-0.5,-0.5);
\draw(0,0.5)       node{\phantom{\textbullet}};
\draw(0.5,0)        node{\phantom{\textbullet}};
\draw(0,-0.5)        node{\textbullet};
\draw(-0.5,0)      node{\phantom{\textbullet}};
\end{tikzpicture}}}

\DeclareRobustCommand{\lvr}{\resizebox{10pt}{!}{\!\!\!\!
\begin{tikzpicture}[line join=round,x=5pt,y=5pt]   
\draw[black] (-1,0)--(1,0);
\draw[black] (0,-1)--(0,1);
\draw(0,0.45)     node{\phantom{\textbullet}};
\draw(0.45,0)      node{\phantom{\textbullet}};
\draw(0,-0.45)     node{\phantom{\textbullet}};
\draw(-0.45,0)     node{\textbullet};
\end{tikzpicture}}}
\DeclareRobustCommand{\dvr}{\resizebox{10pt}{!}{\!\!\!\!
\begin{tikzpicture}[line join=round,x=5pt,y=5pt]   
\draw[black] (-1,0)--(1,0);
\draw[black] (0,-1)--(0,1);
\draw(0,0.45)     node{\phantom{\textbullet}};
\draw(0.45,0)      node{\phantom{\textbullet}};
\draw(0,-0.45)     node{\textbullet};
\draw(-0.45,0)     node{\phantom{\textbullet}};
\end{tikzpicture}}}
\DeclareRobustCommand{\uvr}{\resizebox{10pt}{!}{\!\!\!\!
\begin{tikzpicture}[line join=round,x=5pt,y=5pt]   
\draw[black] (-1,0)--(1,0);
\draw[black] (0,-1)--(0,1);
\draw(0,0.45)     node{\textbullet};
\draw(0.45,0)      node{\phantom{\textbullet}};
\draw(0,-0.45)     node{\phantom{\textbullet}};
\draw(-0.45,0)     node{\phantom{\textbullet}};
\end{tikzpicture}}}
\DeclareRobustCommand{\rvr}{\resizebox{10pt}{!}{\!\!\!\!
\begin{tikzpicture}[line join=round,x=5pt,y=5pt]   
\draw[black] (-1,0)--(1,0);
\draw[black] (0,-1)--(0,1);
\draw(0,0.45)     node{\phantom{\textbullet}};
\draw(0.45,0)      node{\textbullet};
\draw(0,-0.45)     node{\phantom{\textbullet}};
\draw(-0.45,0)     node{\phantom{\textbullet}};
\end{tikzpicture}}}
\DeclareRobustCommand{\cube}{\resizebox{4pt}{!}{\!\!\!\!
\begin{tikzpicture}[line join=round,x=5pt,y=5pt,z=2.5pt]   
\draw[black](1,1,0)--(0,1,0)--(0,1,1)--(1,1,1)--(1,1,0)--(1,0,0)--(1,0,1)--(0,0,1)--(0,1,1);   
\draw[black](1,1,1)--(1,0,1);   
\draw[black](1,0,0)--(0,0,0)--(0,1,0);   
\draw[black](0,0,0)--(0,0,1);
\end{tikzpicture}}}
\DeclareRobustCommand{\zzz}{\resizebox{10pt}{!}{\!\!\!\!\!
\begin{tikzpicture}[line join=round,x=10pt,y=10pt,z=5pt]   
\draw[black](1,1,0)--(0,1,0)--(0,1,1)--(1,1,1)--(1,1,0)--(1,0,0)--(1,0,1)--(0,0,1)--(0,1,1);   
\draw[black](1,1,1)--(1,0,1);   
\draw[black](1,0,0)--(0,0,0)--(0,1,0);   
\draw[black](0,0,0)--(0,0,1); 
\draw(0,0,0) node{\textbullet}; 
\draw(0,0,0) node{\phantom{\textbullet}};
\draw(0,1,0) node{\phantom{\textbullet}};
\draw(1,0,0) node{\phantom{\textbullet}}; 
\draw(0,0,1) node{\phantom{\textbullet}};
\draw(1,1,0) node{\phantom{\textbullet}}; 
\draw(0,1,1) node{\phantom{\textbullet}};
\draw(1,0,1) node{\phantom{\textbullet}}; 
\draw(1,1,1) node{\phantom{\textbullet}};
\end{tikzpicture}}}
\DeclareRobustCommand{\zzo}{\resizebox{10pt}{!}{\!\!\!\!\!
\begin{tikzpicture}[line join=round,x=10pt,y=10pt,z=5pt]   
\draw[black](1,1,0)--(0,1,0)--(0,1,1)--(1,1,1)--(1,1,0)--(1,0,0)--(1,0,1)--(0,0,1)--(0,1,1);   
\draw[black](1,1,1)--(1,0,1);   
\draw[black](1,0,0)--(0,0,0)--(0,1,0);   
\draw[black](0,0,0)--(0,0,1); \draw(0,0,1) node{\textbullet}; 
\draw(0,0,0) node{\phantom{\textbullet}}; \draw(0,1,0) node{\phantom{\textbullet}};
\draw(1,0,0) node{\phantom{\textbullet}}; \draw(0,0,1) node{\phantom{\textbullet}};
\draw(1,1,0) node{\phantom{\textbullet}}; \draw(0,1,1) node{\phantom{\textbullet}};
\draw(1,0,1) node{\phantom{\textbullet}}; \draw(1,1,1) node{\phantom{\textbullet}};
\end{tikzpicture}}}
\DeclareRobustCommand{\zoz}{\resizebox{10pt}{!}{\!\!\!\!\!
\begin{tikzpicture}[line join=round,x=10pt,y=10pt,z=5pt]   
\draw[black](1,1,0)--(0,1,0)--(0,1,1)--(1,1,1)--(1,1,0)--(1,0,0)--(1,0,1)--(0,0,1)--(0,1,1);   
\draw[black](1,1,1)--(1,0,1);   
\draw[black](1,0,0)--(0,0,0)--(0,1,0);   
\draw[black](0,0,0)--(0,0,1); \draw(0,1,0) node{\textbullet}; 
\draw(0,0,0) node{\phantom{\textbullet}}; \draw(0,1,0) node{\phantom{\textbullet}};
\draw(1,0,0) node{\phantom{\textbullet}}; \draw(0,0,1) node{\phantom{\textbullet}};
\draw(1,1,0) node{\phantom{\textbullet}}; \draw(0,1,1) node{\phantom{\textbullet}};
\draw(1,0,1) node{\phantom{\textbullet}}; \draw(1,1,1) node{\phantom{\textbullet}};
\end{tikzpicture}}}
\DeclareRobustCommand{\ozz}{\resizebox{10pt}{!}{\!\!\!\!\!
\begin{tikzpicture}[line join=round,x=10pt,y=10pt,z=5pt]   
\draw[black](1,1,0)--(0,1,0)--(0,1,1)--(1,1,1)--(1,1,0)--(1,0,0)--(1,0,1)--(0,0,1)--(0,1,1);   
\draw[black](1,1,1)--(1,0,1);   
\draw[black](1,0,0)--(0,0,0)--(0,1,0);   
\draw[black](0,0,0)--(0,0,1); \draw(1,0,0) node{\textbullet}; 
\draw(0,0,0) node{\phantom{\textbullet}}; \draw(0,1,0) node{\phantom{\textbullet}};
\draw(1,0,0) node{\phantom{\textbullet}}; \draw(0,0,1) node{\phantom{\textbullet}};
\draw(1,1,0) node{\phantom{\textbullet}}; \draw(0,1,1) node{\phantom{\textbullet}};
\draw(1,0,1) node{\phantom{\textbullet}}; \draw(1,1,1) node{\phantom{\textbullet}};
\end{tikzpicture}}}
\DeclareRobustCommand{\zoo}{\resizebox{10pt}{!}{\!\!\!\!\!
\begin{tikzpicture}[line join=round,x=10pt,y=10pt,z=5pt]   
\draw[black](1,1,0)--(0,1,0)--(0,1,1)--(1,1,1)--(1,1,0)--(1,0,0)--(1,0,1)--(0,0,1)--(0,1,1);   
\draw[black](1,1,1)--(1,0,1);   
\draw[black](1,0,0)--(0,0,0)--(0,1,0);   
\draw[black](0,0,0)--(0,0,1); \draw(0,1,1) node{\textbullet}; 
\draw(0,0,0) node{\phantom{\textbullet}}; \draw(0,1,0) node{\phantom{\textbullet}};
\draw(1,0,0) node{\phantom{\textbullet}}; \draw(0,0,1) node{\phantom{\textbullet}};
\draw(1,1,0) node{\phantom{\textbullet}}; \draw(0,1,1) node{\phantom{\textbullet}};
\draw(1,0,1) node{\phantom{\textbullet}}; \draw(1,1,1) node{\phantom{\textbullet}};
\end{tikzpicture}}}
\DeclareRobustCommand{\ozo}{\resizebox{10pt}{!}{\!\!\!\!\!
\begin{tikzpicture}[line join=round,x=10pt,y=10pt,z=5pt]   
\draw[black](1,1,0)--(0,1,0)--(0,1,1)--(1,1,1)--(1,1,0)--(1,0,0)--(1,0,1)--(0,0,1)--(0,1,1);   
\draw[black](1,1,1)--(1,0,1);   
\draw[black](1,0,0)--(0,0,0)--(0,1,0);   
\draw[black](0,0,0)--(0,0,1); \draw(1,0,1) node{\textbullet}; 
\draw(0,0,0) node{\phantom{\textbullet}}; \draw(0,1,0) node{\phantom{\textbullet}};
\draw(1,0,0) node{\phantom{\textbullet}}; \draw(0,0,1) node{\phantom{\textbullet}};
\draw(1,1,0) node{\phantom{\textbullet}}; \draw(0,1,1) node{\phantom{\textbullet}};
\draw(1,0,1) node{\phantom{\textbullet}}; \draw(1,1,1) node{\phantom{\textbullet}};
\end{tikzpicture}}}
\DeclareRobustCommand{\ooz}{\resizebox{10pt}{!}{\!\!\!\!\!
\begin{tikzpicture}[line join=round,x=10pt,y=10pt,z=5pt]   
\draw[black](1,1,0)--(0,1,0)--(0,1,1)--(1,1,1)--(1,1,0)--(1,0,0)--(1,0,1)--(0,0,1)--(0,1,1);   
\draw[black](1,1,1)--(1,0,1);   
\draw[black](1,0,0)--(0,0,0)--(0,1,0);   
\draw[black](0,0,0)--(0,0,1); \draw(1,1,0) node{\textbullet}; 
\draw(0,0,0) node{\phantom{\textbullet}}; \draw(0,1,0) node{\phantom{\textbullet}};
\draw(1,0,0) node{\phantom{\textbullet}}; \draw(0,0,1) node{\phantom{\textbullet}};
\draw(1,1,0) node{\phantom{\textbullet}}; \draw(0,1,1) node{\phantom{\textbullet}};
\draw(1,0,1) node{\phantom{\textbullet}}; \draw(1,1,1) node{\phantom{\textbullet}};
\end{tikzpicture}}}
\DeclareRobustCommand{\ooo}{\resizebox{10pt}{!}{\!\!\!\!\!
\begin{tikzpicture}[line join=round,x=10pt,y=10pt,z=5pt]   
\draw[black](1,1,0)--(0,1,0)--(0,1,1)--(1,1,1)--(1,1,0)--(1,0,0)--(1,0,1)--(0,0,1)--(0,1,1);   
\draw[black](1,1,1)--(1,0,1);   
\draw[black](1,0,0)--(0,0,0)--(0,1,0);   
\draw[black](0,0,0)--(0,0,1); \draw(1,1,1) node{\textbullet}; 
\draw(0,0,0) node{\phantom{\textbullet}}; \draw(0,1,0) node{\phantom{\textbullet}};
\draw(1,0,0) node{\phantom{\textbullet}}; \draw(0,0,1) node{\phantom{\textbullet}};
\draw(1,1,0) node{\phantom{\textbullet}}; \draw(0,1,1) node{\phantom{\textbullet}};
\draw(1,0,1) node{\phantom{\textbullet}}; \draw(1,1,1) node{\phantom{\textbullet}};
\end{tikzpicture}}}
\DeclareRobustCommand{\tzzz}{\resizebox{10pt}{!}{\!\!\!\!\!
\begin{tikzpicture}[line join=round,x=10pt,y=10pt,z=5pt]   
\draw[black](1,1,0)--(0,1,0)--(0,1,1)--(1,1,1)--(1,1,0)--(1,0,0)--(1,0,1)--(0,0,1)--(0,1,1);   
\draw[black](1,1,1)--(1,0,1);   
\draw[black](1,0,0)--(0,0,0)--(0,1,0);   
\draw[black](0,0,0)--(0,0,1); \draw(0,0,0) node{$\times$}; 
\draw(0,0,0) node{\phantom{\textbullet}}; \draw(0,1,0) node{\phantom{\textbullet}};
\draw(1,0,0) node{\phantom{\textbullet}}; \draw(0,0,1) node{\phantom{\textbullet}};
\draw(1,1,0) node{\phantom{\textbullet}}; \draw(0,1,1) node{\phantom{\textbullet}};
\draw(1,0,1) node{\phantom{\textbullet}}; \draw(1,1,1) node{\phantom{\textbullet}};
\end{tikzpicture}}}
\DeclareRobustCommand{\tzzo}{\resizebox{10pt}{!}{\!\!\!\!\!
\begin{tikzpicture}[line join=round,x=10pt,y=10pt,z=5pt]   
\draw[black](1,1,0)--(0,1,0)--(0,1,1)--(1,1,1)--(1,1,0)--(1,0,0)--(1,0,1)--(0,0,1)--(0,1,1);   
\draw[black](1,1,1)--(1,0,1);   
\draw[black](1,0,0)--(0,0,0)--(0,1,0);   
\draw[black](0,0,0)--(0,0,1); \draw(0,0,1) node{$\times$}; 
\draw(0,0,0) node{\phantom{\textbullet}}; \draw(0,1,0) node{\phantom{\textbullet}};
\draw(1,0,0) node{\phantom{\textbullet}}; \draw(0,0,1) node{\phantom{\textbullet}};
\draw(1,1,0) node{\phantom{\textbullet}}; \draw(0,1,1) node{\phantom{\textbullet}};
\draw(1,0,1) node{\phantom{\textbullet}}; \draw(1,1,1) node{\phantom{\textbullet}};
\end{tikzpicture}}}
\DeclareRobustCommand{\tzoz}{\resizebox{10pt}{!}{\!\!\!\!\!
\begin{tikzpicture}[line join=round,x=10pt,y=10pt,z=5pt]   
\draw[black](1,1,0)--(0,1,0)--(0,1,1)--(1,1,1)--(1,1,0)--(1,0,0)--(1,0,1)--(0,0,1)--(0,1,1);   
\draw[black](1,1,1)--(1,0,1);   
\draw[black](1,0,0)--(0,0,0)--(0,1,0);   
\draw[black](0,0,0)--(0,0,1); \draw(0,1,0) node{$\times$}; 
\draw(0,0,0) node{\phantom{\textbullet}}; \draw(0,1,0) node{\phantom{\textbullet}};
\draw(1,0,0) node{\phantom{\textbullet}}; \draw(0,0,1) node{\phantom{\textbullet}};
\draw(1,1,0) node{\phantom{\textbullet}}; \draw(0,1,1) node{\phantom{\textbullet}};
\draw(1,0,1) node{\phantom{\textbullet}}; \draw(1,1,1) node{\phantom{\textbullet}};
\end{tikzpicture}}}
\DeclareRobustCommand{\tozz}{\resizebox{10pt}{!}{\!\!\!\!\!
\begin{tikzpicture}[line join=round,x=10pt,y=10pt,z=5pt]   
\draw[black](1,1,0)--(0,1,0)--(0,1,1)--(1,1,1)--(1,1,0)--(1,0,0)--(1,0,1)--(0,0,1)--(0,1,1);   
\draw[black](1,1,1)--(1,0,1);   
\draw[black](1,0,0)--(0,0,0)--(0,1,0);   
\draw[black](0,0,0)--(0,0,1); \draw(1,0,0) node{$\times$}; 
\draw(0,0,0) node{\phantom{\textbullet}}; \draw(0,1,0) node{\phantom{\textbullet}};
\draw(1,0,0) node{\phantom{\textbullet}}; \draw(0,0,1) node{\phantom{\textbullet}};
\draw(1,1,0) node{\phantom{\textbullet}}; \draw(0,1,1) node{\phantom{\textbullet}};
\draw(1,0,1) node{\phantom{\textbullet}}; \draw(1,1,1) node{\phantom{\textbullet}};
\end{tikzpicture}}}
\DeclareRobustCommand{\tzoo}{\resizebox{10pt}{!}{\!\!\!\!\!
\begin{tikzpicture}[line join=round,x=10pt,y=10pt,z=5pt]   
\draw[black](1,1,0)--(0,1,0)--(0,1,1)--(1,1,1)--(1,1,0)--(1,0,0)--(1,0,1)--(0,0,1)--(0,1,1);   
\draw[black](1,1,1)--(1,0,1);   
\draw[black](1,0,0)--(0,0,0)--(0,1,0);   
\draw[black](0,0,0)--(0,0,1); \draw(0,1,1) node{$\times$}; 
\draw(0,0,0) node{\phantom{\textbullet}}; \draw(0,1,0) node{\phantom{\textbullet}};
\draw(1,0,0) node{\phantom{\textbullet}}; \draw(0,0,1) node{\phantom{\textbullet}};
\draw(1,1,0) node{\phantom{\textbullet}}; \draw(0,1,1) node{\phantom{\textbullet}};
\draw(1,0,1) node{\phantom{\textbullet}}; \draw(1,1,1) node{\phantom{\textbullet}};
\end{tikzpicture}}}
\DeclareRobustCommand{\tozo}{\resizebox{10pt}{!}{\!\!\!\!\!
\begin{tikzpicture}[line join=round,x=10pt,y=10pt,z=5pt]   
\draw[black](1,1,0)--(0,1,0)--(0,1,1)--(1,1,1)--(1,1,0)--(1,0,0)--(1,0,1)--(0,0,1)--(0,1,1);   
\draw[black](1,1,1)--(1,0,1);   
\draw[black](1,0,0)--(0,0,0)--(0,1,0);   
\draw[black](0,0,0)--(0,0,1); \draw(1,0,1) node{$\times$}; 
\draw(0,0,0) node{\phantom{\textbullet}}; \draw(0,1,0) node{\phantom{\textbullet}};
\draw(1,0,0) node{\phantom{\textbullet}}; \draw(0,0,1) node{\phantom{\textbullet}};
\draw(1,1,0) node{\phantom{\textbullet}}; \draw(0,1,1) node{\phantom{\textbullet}};
\draw(1,0,1) node{\phantom{\textbullet}}; \draw(1,1,1) node{\phantom{\textbullet}};
\end{tikzpicture}}}
\DeclareRobustCommand{\tooz}{\resizebox{10pt}{!}{\!\!\!\!\!
\begin{tikzpicture}[line join=round,x=10pt,y=10pt,z=5pt]   
\draw[black](1,1,0)--(0,1,0)--(0,1,1)--(1,1,1)--(1,1,0)--(1,0,0)--(1,0,1)--(0,0,1)--(0,1,1);   
\draw[black](1,1,1)--(1,0,1);   
\draw[black](1,0,0)--(0,0,0)--(0,1,0);   
\draw[black](0,0,0)--(0,0,1); \draw(1,1,0) node{$\times$}; 
\draw(0,0,0) node{\phantom{\textbullet}}; \draw(0,1,0) node{\phantom{\textbullet}};
\draw(1,0,0) node{\phantom{\textbullet}}; \draw(0,0,1) node{\phantom{\textbullet}};
\draw(1,1,0) node{\phantom{\textbullet}}; \draw(0,1,1) node{\phantom{\textbullet}};
\draw(1,0,1) node{\phantom{\textbullet}}; \draw(1,1,1) node{\phantom{\textbullet}};
\end{tikzpicture}}}
\DeclareRobustCommand{\tooo}{\resizebox{10pt}{!}{\!\!\!\!\!
\begin{tikzpicture}[line join=round,x=10pt,y=10pt,z=5pt]   
\draw[black](1,1,0)--(0,1,0)--(0,1,1)--(1,1,1)--(1,1,0)--(1,0,0)--(1,0,1)--(0,0,1)--(0,1,1);   
\draw[black](1,1,1)--(1,0,1);   
\draw[black](1,0,0)--(0,0,0)--(0,1,0);   
\draw[black](0,0,0)--(0,0,1); \draw(1,1,1) node{$\times$}; 
\draw(0,0,0) node{\phantom{\textbullet}}; \draw(0,1,0) node{\phantom{\textbullet}};
\draw(1,0,0) node{\phantom{\textbullet}}; \draw(0,0,1) node{\phantom{\textbullet}};
\draw(1,1,0) node{\phantom{\textbullet}}; \draw(0,1,1) node{\phantom{\textbullet}};
\draw(1,0,1) node{\phantom{\textbullet}}; \draw(1,1,1) node{\phantom{\textbullet}};
\end{tikzpicture}}}
\DeclareRobustCommand{\sdots}{\resizebox{10pt}{!}{\!\!\!\!\!
\begin{tikzpicture}[line join=round,x=10pt,y=10pt,z=5pt]   
\draw[black](1,1,0)--(0,1,0)--(0,1,1)--(1,1,1)--(1,1,0)--(1,0,0)--(1,0,1)--(0,0,1)--(0,1,1);   
\draw[black](1,1,1)--(1,0,1);   
\draw[black](1,0,0)--(0,0,0)--(0,1,0);   
\draw[black](0,0,0)--(0,0,1);
\draw(0,0,0) node{\phantom{\textbullet}}; \draw(0,1,0) node{\textbullet};
\draw(1,0,0) node{\textbullet}; \draw(0,0,1) node{\phantom{\textbullet}};
\draw(1,1,0) node{\textbullet}; \draw(0,1,1) node{\phantom{\textbullet}};
\draw(1,0,1) node{\phantom{\textbullet}}; \draw(1,1,1) node{\textbullet};
\end{tikzpicture}}}
\DeclareRobustCommand{\mdots}{\resizebox{10pt}{!}{\!\!\!\!\!
\begin{tikzpicture}[line join=round,x=10pt,y=10pt,z=5pt]   
\draw[black](1,1,0)--(0,1,0)--(0,1,1)--(1,1,1)--(1,1,0)--(1,0,0)--(1,0,1)--(0,0,1)--(0,1,1);   
\draw[black](1,1,1)--(1,0,1);   
\draw[black](1,0,0)--(0,0,0)--(0,1,0);   
\draw[black](0,0,0)--(0,0,1);
\draw(0,0,0) node{\phantom{\textbullet}}; \draw(0,1,0) node{\textbullet};
\draw(1,0,0) node{\textbullet}; \draw(0,0,1) node{\textbullet};
\draw(1,1,0) node{\phantom{\textbullet}}; \draw(0,1,1) node{\phantom{\textbullet}};
\draw(1,0,1) node{\phantom{\textbullet}}; \draw(1,1,1) node{\textbullet};
\end{tikzpicture}}}
\DeclareRobustCommand{\scros}{\resizebox{10pt}{!}{\!\!\!\!\!
\begin{tikzpicture}[line join=round,x=10pt,y=10pt,z=5pt]   
\draw[black](1,1,0)--(0,1,0)--(0,1,1)--(1,1,1)--(1,1,0)--(1,0,0)--(1,0,1)--(0,0,1)--(0,1,1);   
\draw[black](1,1,1)--(1,0,1);   
\draw[black](1,0,0)--(0,0,0)--(0,1,0);   
\draw[black](0,0,0)--(0,0,1);
\draw(0,0,0) node{$\times$}; \draw(0,1,0) node{\phantom{\textbullet}};
\draw(1,0,0) node{\phantom{\textbullet}}; \draw(0,0,1) node{\phantom{\textbullet}};
\draw(1,1,0) node{$\times$}; \draw(0,1,1) node{$\times$};
\draw(1,0,1) node{$\times$}; \draw(1,1,1) node{\phantom{\textbullet}};
\end{tikzpicture}}}
\DeclareRobustCommand{\mcros}{\resizebox{10pt}{!}{\!\!\!\!\!
\begin{tikzpicture}[line join=round,x=10pt,y=10pt,z=5pt]   
\draw[black](1,1,1)--(1,0,1);   
\draw[black](1,1,0)--(0,1,0)--(0,1,1)--(1,1,1)--(1,1,0)--(1,0,0)--(1,0,1)--(0,0,1)--(0,1,1);   
\draw[black](1,0,0)--(0,0,0)--(0,1,0);   
\draw[black](0,0,0)--(0,0,1);
\draw(0,0,0) node{$\times$}; \draw(0,1,0) node{\phantom{\textbullet}};
\draw(1,0,0) node{\phantom{\textbullet}}; \draw(0,0,1) node{$\times$};
\draw(1,1,0) node{\phantom{\textbullet}}; \draw(0,1,1) node{$\times$};
\draw(1,0,1) node{$\times$}; \draw(1,1,1) node{\phantom{\textbullet}};
\end{tikzpicture}}}
\DeclareRobustCommand{\hex}{\resizebox{3.5pt}{!}{\!\!\!\!
\begin{tikzpicture}[line join=round,x=5pt,y=5pt]   
\draw[black] (0,0)--(0.866,0.5)--(0.866,1.5)--(0,2)--(-0.866,1.5)--(-0.866,0.5)--(0,0);
\end{tikzpicture}}}
\DeclareRobustCommand{\hone}{\resizebox{10pt}{!}{\!\!\!\!
\begin{tikzpicture}[line join=round,x=5pt,y=5pt]   
\draw[black] (0,0)--(0.866,0.5)--(0.866,1.5)--(0,2)--(-0.866,1.5)--(-0.866,0.5)--(0,0);
\draw(0,0)        node{\phantom{\textbullet}};
\draw(-0.866,0.5) node{\textbullet};
\draw(-0.866,1.5) node{\phantom{\textbullet}};
\draw(0,2)        node{\phantom{\textbullet}};
\draw(0.866,1.5)  node{\phantom{\textbullet}};
\draw(0.866,0.5)  node{\phantom{\textbullet}};
\end{tikzpicture}}}
\DeclareRobustCommand{\htwo}{\resizebox{10pt}{!}{\!\!\!\!
\begin{tikzpicture}[line join=round,x=5pt,y=5pt]   
\draw[black] (0,0)--(0.866,0.5)--(0.866,1.5)--(0,2)--(-0.866,1.5)--(-0.866,0.5)--(0,0);
\draw(0,0)        node{\phantom{\textbullet}};
\draw(-0.866,0.5) node{\phantom{\textbullet}};
\draw(-0.866,1.5) node{\textbullet};
\draw(0,2)        node{\phantom{\textbullet}};
\draw(0.866,1.5)  node{\phantom{\textbullet}};
\draw(0.866,0.5)  node{\phantom{\textbullet}};
\end{tikzpicture}}}
\DeclareRobustCommand{\hthr}{\resizebox{10pt}{!}{\!\!\!\!
\begin{tikzpicture}[line join=round,x=5pt,y=5pt]   
\draw[black] (0,0)--(0.866,0.5)--(0.866,1.5)--(0,2)--(-0.866,1.5)--(-0.866,0.5)--(0,0);
\draw(0,0)        node{\phantom{\textbullet}};
\draw(-0.866,0.5) node{\phantom{\textbullet}};
\draw(-0.866,1.5) node{\phantom{\textbullet}};
\draw(0,2)        node{\textbullet};
\draw(0.866,1.5)  node{\phantom{\textbullet}};
\draw(0.866,0.5)  node{\phantom{\textbullet}};
\end{tikzpicture}}}
\DeclareRobustCommand{\hfou}{\resizebox{10pt}{!}{\!\!\!\!
\begin{tikzpicture}[line join=round,x=5pt,y=5pt]   
\draw[black] (0,0)--(0.866,0.5)--(0.866,1.5)--(0,2)--(-0.866,1.5)--(-0.866,0.5)--(0,0);
\draw(0,0)        node{\phantom{\textbullet}};
\draw(-0.866,0.5) node{\phantom{\textbullet}};
\draw(-0.866,1.5) node{\phantom{\textbullet}};
\draw(0,2)        node{\phantom{\textbullet}};
\draw(0.866,1.5)  node{\textbullet};
\draw(0.866,0.5)  node{\phantom{\textbullet}};
\end{tikzpicture}}}
\DeclareRobustCommand{\hfiv}{\resizebox{10pt}{!}{\!\!\!\!
\begin{tikzpicture}[line join=round,x=5pt,y=5pt]   
\draw[black] (0,0)--(0.866,0.5)--(0.866,1.5)--(0,2)--(-0.866,1.5)--(-0.866,0.5)--(0,0);
\draw(0,0)        node{\phantom{\textbullet}};
\draw(-0.866,0.5) node{\phantom{\textbullet}};
\draw(-0.866,1.5) node{\phantom{\textbullet}};
\draw(0,2)        node{\phantom{\textbullet}};
\draw(0.866,1.5)  node{\phantom{\textbullet}};
\draw(0.866,0.5)  node{\textbullet};
\end{tikzpicture}}}
\DeclareRobustCommand{\hsix}{\resizebox{10pt}{!}{\!\!\!\!
\begin{tikzpicture}[line join=round,x=5pt,y=5pt]   
\draw[black] (0,0)--(0.866,0.5)--(0.866,1.5)--(0,2)--(-0.866,1.5)--(-0.866,0.5)--(0,0);
\draw(0,0)        node{\textbullet};
\draw(-0.866,0.5) node{\phantom{\textbullet}};
\draw(-0.866,1.5) node{\phantom{\textbullet}};
\draw(0,2)        node{\phantom{\textbullet}};
\draw(0.866,1.5)  node{\phantom{\textbullet}};
\draw(0.866,0.5)  node{\phantom{\textbullet}};
\end{tikzpicture}}}
\DeclareRobustCommand{\symx}{\rotatebox[origin=c]{60}{$\leftrightsquigarrow$}}

\title{General phase spaces: from discrete variables to rotor and continuum
limits}

\author{Victor~V.~Albert}
\email{valbert4@gmail.com}

\address{Yale Quantum Institute, Yale University, New Haven, Connecticut 06520,
USA}

\author{Saverio Pascazio}

\affiliation{Dipartimento di Fisica and MECENAS, Università di Bari, I-70126 Bari,
Italy~\\
Istituto Nazionale di Ottica (INO-CNR), I-50125 Firenze, Italy~\\
INFN, Sezione di Bari, I-70126 Bari, Italy}

\author{Michel~H.~Devoret}
\email{michel.devoret@yale.edu}

\address{Department of Applied Physics, Yale University, New Haven, Connecticut
06520, USA}
\begin{abstract}
We provide a basic introduction to discrete-variable, rotor, and continuous-variable
quantum phase spaces, explaining how the latter two can be understood
as limiting cases of the first. We extend the limit-taking procedures
used to travel between phase spaces to a general class of Hamiltonians
(including many local stabilizer codes) and provide six examples:
the Harper equation, the Baxter parafermionic spin chain, the Rabi
model, the Kitaev toric code, the Haah cubic code (which we generalize
to qudits), and the Kitaev honeycomb model. We obtain continuous-variable
generalizations of all models, some of which are novel. The Baxter
model is mapped to a chain of coupled oscillators and the Rabi model
to the optomechanical radiation pressure Hamiltonian. The procedures
also yield rotor versions of all models, five of which are novel many-body
extensions of the almost Mathieu equation. The toric and cubic codes
are mapped to lattice models of rotors, with the toric code case related
to $U(1)$ lattice gauge theory.
\end{abstract}


\maketitle

\section{Introduction}

Continuous-variable ($\ft$) limits of discrete-variable ($\dfs$)
systems, if done carefully, can yield new models which are both interesting
and helpful in illuminating low-energy properties of the original
$\dfs$ systems. A physical example comes from spin-wave theory, where
the Hamiltonian of interacting spins $S$ is expanded in the limit
of small quantum fluctuations ($S\gg1$) using the Holstein-Primakoff
mapping \cite{auerbackbook}. Another type of limit involves thinking
of each $\dfs$ system not as a spin, but as a finite quantum system
\cite{Vourdas2017} of dimension $\N=2S+1$ (in quantum information,
a qu$\N$it) whose conjugate variables, position and momentum, are
bounded and discrete. This $\dfs\rightarrow\ft$ limit then involves
making both variables continuous and unbounded. A less-used version
makes one of the variables continuous and periodic (i.e., an angle)
and the other an integer, resulting in the phase space of a rotor
($\dfs\rightarrow\fs$; see Table \ref{t:1} for all three phase spaces).
While these limits deal with the underlying phase space of a $\dfs$
system, it is not always clear when and how to apply them to specific
$\dfs$ Hamiltonians, particularly in composite scenarios (e.g., the
Jaynes Cummings model) consisting of both $\dfs$ and $\ft$ components.
We attempt to address these issues by outlining general and straightforward
limit-taking procedures and applying them to obtain known and new
$\dfs$, $\fs$, and $\ft$ extensions of six well-known models from
condensed-matter physics and quantum computation (see Table \ref{t:2}).

In Sec. \ref{sec:tps}, we provide a basic introduction of the three
aforementioned phase spaces ($\dfs$, $\fs$, and $\ft$) such that
the latter two can be thought of as limits of the former. In turn,
$\dfs$ phase space can be understood as a discretization of $\ft$
phase space in terms of fixed position and momentum increments $\delta x$
and $\delta p$ (similar to a computer approximating differential
equations with difference equations). Section \ref{sec:Continuum-and-rotor}
outlines all limit-taking procedures, with comments on when $\dfs\rightarrow\ft$
can be a valid low-energy approximation of a $\dfs$ Hamiltonian.
We warm up with a known and exactly-solvable example of all limit-taking
procedures \textemdash{} the harmonic oscillator AKA the Harper equation
\textemdash{} in Sec. \ref{sec:sho}. In Sec. \ref{sec:Baxter}, we
study a many-body coupled-oscillator example \textemdash{} the Baxter
$\Z_{\N}$ parafermionic spin chain. In Sec. \ref{sec:rabi2}, we
introduce the Rabi model, show that its $\N$-state extension has
a dihedral symmetry, and provide its analogues in all three phase
spaces. We continue with deriving the $\ft$ toric code model from
the $\dfs$ one while introducing novel rotor toric code extensions
in Sec. \ref{sec:toric-code}. In Sec. \ref{sec:Haah-cubic-code},
we develop $\dfs$, $\fs$, and $\ft$ extensions of the Haah cubic
code. The Kitaev honeycomb model is generalized in Sec. \ref{sec:Kitaev-model}.
A final discussion is given in Sec. \ref{sec:d}.

\section{Classical and quantum phase spaces\label{sec:tps}}

\begin{table*}[t]
\begin{tabular}[t]{>{\raggedright}m{0.8in}ccc}
\toprule 
 & $\dfs$ & $\fs$ & $\ft$\tabularnewline
\midrule 
Sketch & \includegraphics[width=1.3in]{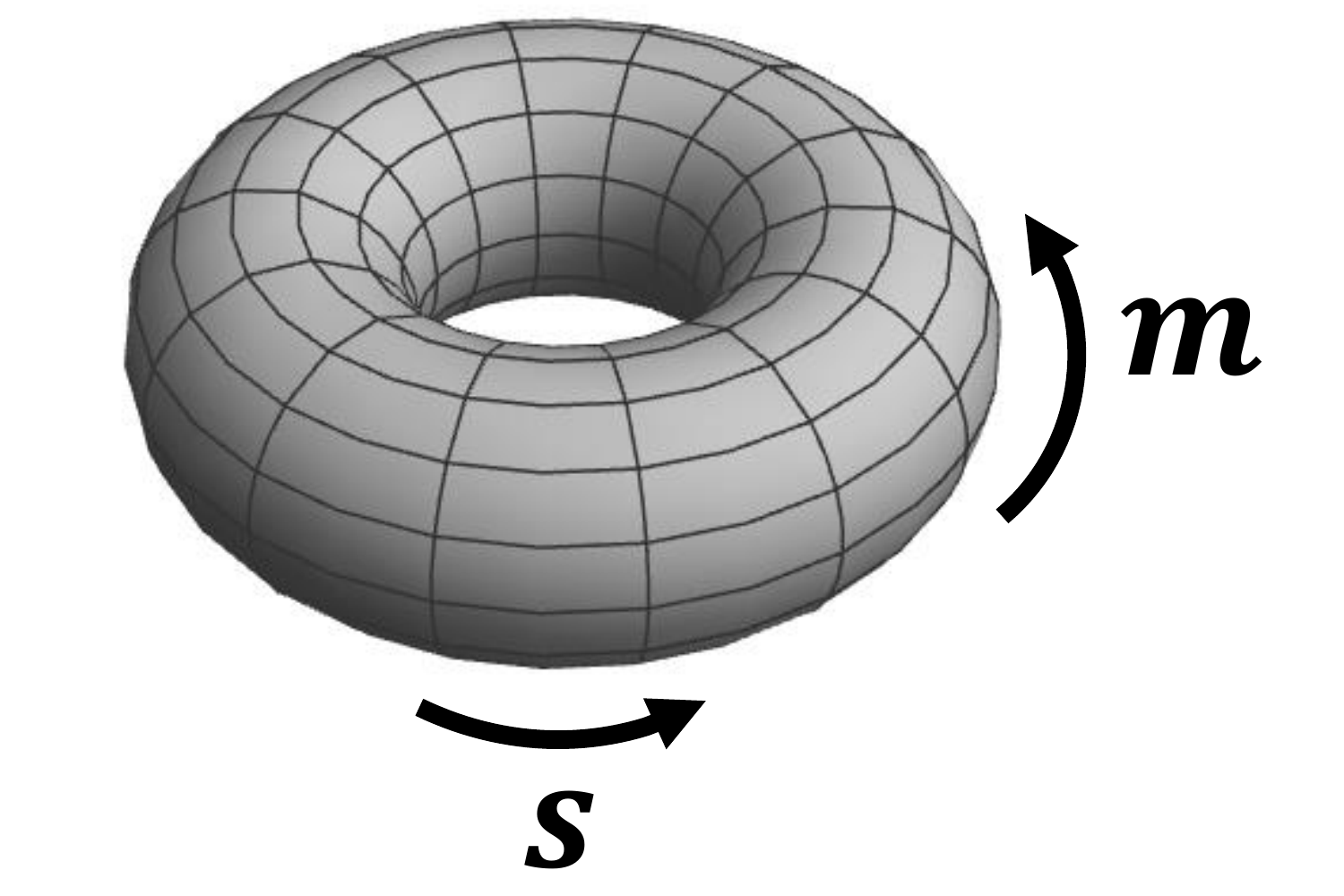} & \includegraphics[width=1.3in]{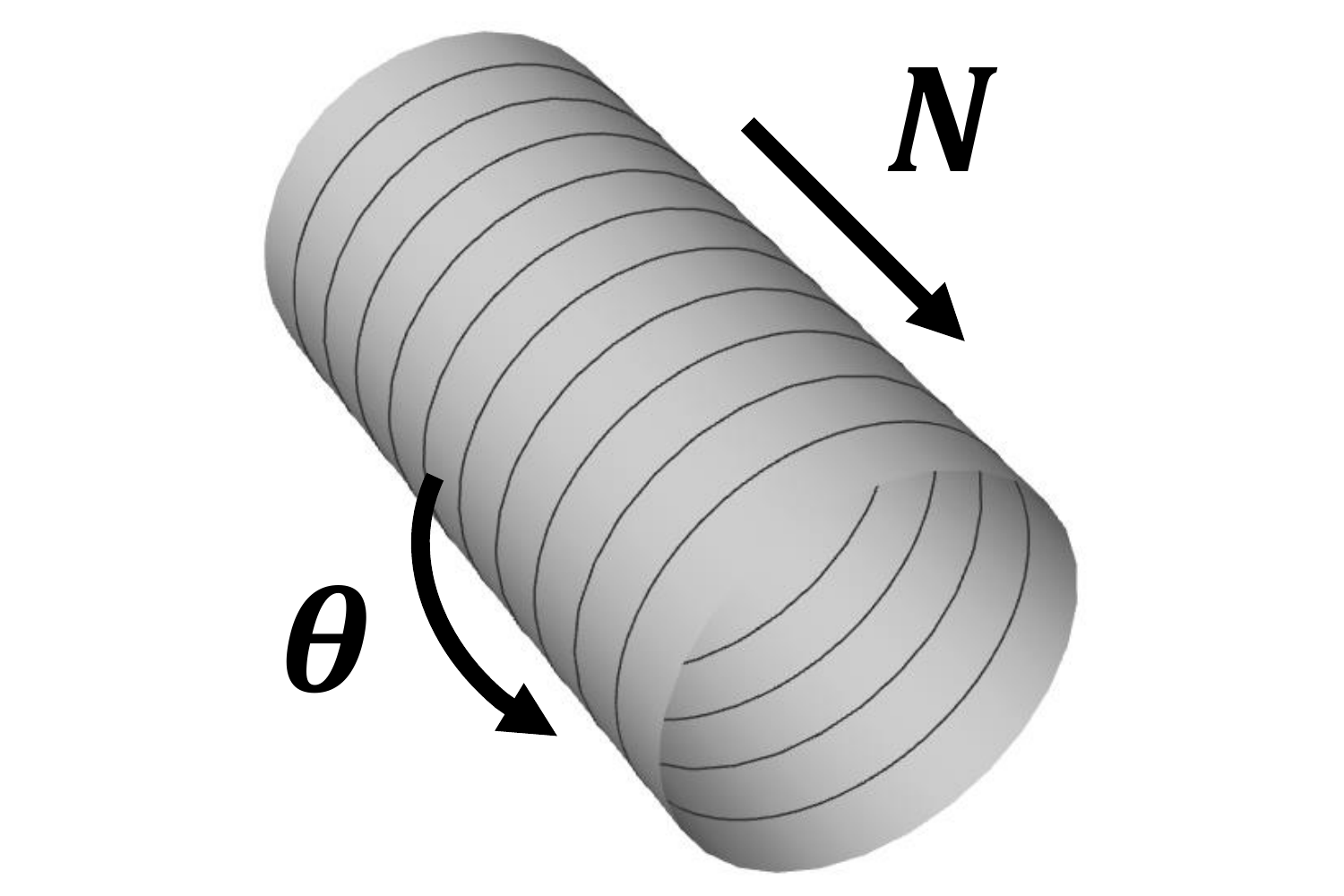} & \includegraphics[width=1.3in]{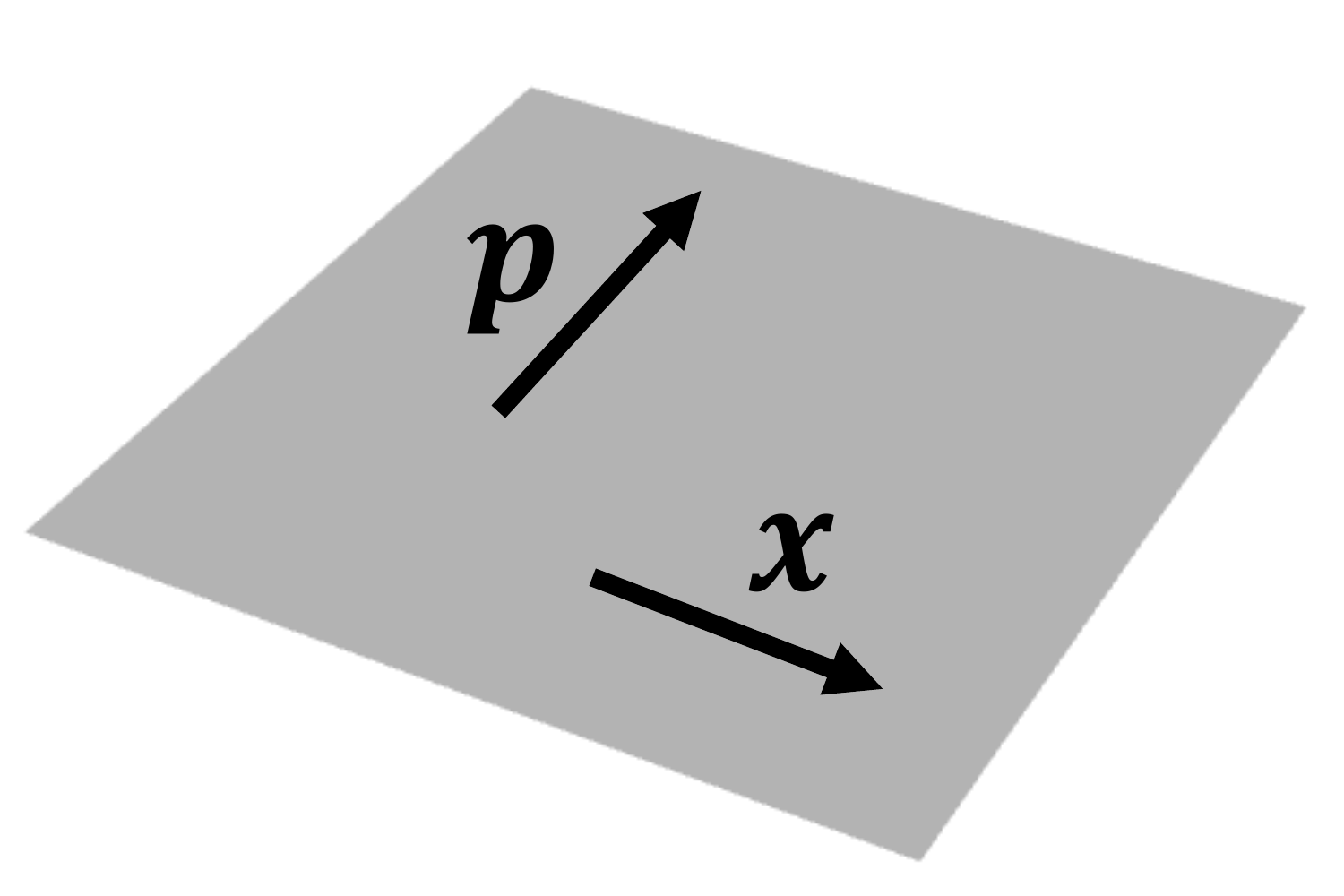}\tabularnewline
\midrule 
Phase space & $(\ci,\ri)\in\Z_{\N}\times\Z_{\N}$ & $(e^{i\theta},N)\in\mathbb{T}\times\Z$ & $(x,p)\in\mathbb{R}\times\mathbb{R}$\tabularnewline
\midrule 
\addlinespace
Weyl relation & ~~~$e^{-i\frac{2\pi}{\N}\r}e^{i\frac{2\pi}{\N}\c}e^{i\frac{2\pi}{\N}\r}e^{-i\frac{2\pi}{\N}\c}=e^{-i\frac{2\pi}{\N}}$~~~ & $e^{-i\NN}e^{i\k}e^{i\NN}e^{-i\k}=e^{-i}$ & $e^{-i\p}e^{i\x}e^{i\p}e^{-i\x}=e^{-i\hbar}$\tabularnewline
\midrule 
CCR & `` $\left[\c,\r\right]=i{\displaystyle \frac{\N}{2\pi}}$~'' & ``~$[\k,\NN]=i$~'' & $\left[\x,\p\right]=i\hbar$\tabularnewline\addlinespace
\midrule 
Fourier & Discrete Fourier transform & Fourier series & Fourier transform\tabularnewline
\midrule 
\multirow{2}{0.8in}{``Bases''} & $G\left(\c\right)|\ci\ket=G\left(\ci\right)|\ci\ket$  & $K(\k)|\theta\ket=K\left(\theta\right)|\theta\ket$ & $\x|x\ket=x|x\ket$\tabularnewline
 & $G\left(\r\right)|\ri\ket=G\left(\ri\right)|\ri\ket$  & $\NN|N\ket=N|N\ket$ & $\p|p\ket=p|p\ket$\tabularnewline
\midrule 
\multirow{2}{0.8in}{Translations} & $e^{-i\frac{2\pi}{\N}\M\r}|\ci\ket=|\ci+\M\ket$ & $e^{-i\vartheta\NN}|\theta\ket=|\theta+\vartheta\ket$ & $e^{-i\xi\p}|x\ket=|x+\xi\ket$\tabularnewline
 & $e^{i\frac{2\pi}{\N}\M\c}|\ri\ket=|\ri+\M\ket$ & $e^{i\M\k}|N\ket=|N+\M\ket$ & $e^{i\xi\x}|p\ket=|p+\xi\ket$\tabularnewline
\midrule 
Character & $\bra\ci|\ri\ket=\frac{1}{\sqrt{\N}}e^{i\frac{2\pi}{\N}\ci\ri}$ & $\bra\theta|N\ket=\frac{1}{\sqrt{2\pi}}e^{i\theta N}$ & $\bra x|p\ket=\frac{1}{\sqrt{2\pi}}e^{ixp}$\tabularnewline
\midrule 
\multirow{2}{0.8in}{Relationship between bases} & $|\ri\ket=\frac{1}{\sqrt{\N}}{\displaystyle \sum_{\ci\in\Z_{\N}}}e^{i\frac{2\pi}{\N}\ci\ri}|\ci\ket$ & $|N\ket=\frac{1}{\sqrt{2\pi}}{\displaystyle \int_{-\pi}^{\pi}}d\theta e^{i\theta N}|\theta\ket$ & $|p\ket=\frac{1}{\sqrt{2\pi}}{\displaystyle \int_{-\infty}^{\infty}}dxe^{ixp}|x\ket$\tabularnewline
 & $|\ci\ket=\frac{1}{\sqrt{\N}}{\displaystyle \sum_{\ri\in\Z_{\N}}}e^{-i\frac{2\pi}{\N}\ci\ri}|\ri\ket$ & ${\displaystyle |\theta\ket=}\frac{1}{\sqrt{2\pi}}{\displaystyle \sum_{N\in\Z}e^{-i\theta N}|N\ket}$ & $|x\ket=\frac{1}{\sqrt{2\pi}}{\displaystyle \int_{-\infty}^{\infty}}dpe^{-ixp}|p\ket$\tabularnewline
\midrule
\multirow{1}{0.8in}{Completeness} & ${\displaystyle \sum_{\ci\in\Z_{\N}}}|\ci\ket\bra\ci|={\displaystyle \sum_{\ri\in\Z_{\N}}}|\ri\ket\bra\ri|=\id$ & ~~~${\displaystyle \sum_{N\in\Z}|N\ket\bra N|={\displaystyle \int_{-\pi}^{\pi}}\frac{d\theta}{2\pi}|\theta\ket\bra\theta|=\id}$~~~ & ~~~${\displaystyle \int_{-\infty}^{\infty}}dx|x\ket\bra x|={\displaystyle \int_{-\infty}^{\infty}}dp|p\ket\bra p|=\id$~~~\tabularnewline
\midrule 
\multirow{2}{0.8in}{Orthonormality} & ${\displaystyle \sum_{\ci\in\Z_{\N}}\frac{e^{i\frac{2\pi}{\N}\left(\ri-\ri'\right)\ci}}{\N}=\delta_{\ri,\ri^{\pr}}}$ & ${\displaystyle {\displaystyle \int_{-\pi}^{\pi}}\frac{d\theta}{2\pi}e^{i\left(N-N'\right)\theta}=\delta_{N,N'}}$ & ${\displaystyle \int_{-\infty}^{\infty}\frac{dx}{2\pi}e^{i\left(p-p'\right)x}=\delta\left(p-p'\right)}$\tabularnewline
 & ${\displaystyle \sum_{\ri\in\Z_{\N}}\frac{e^{i\frac{2\pi}{\N}\left(\ci-\ci'\right)\ri}}{\N}=\delta_{\ci,\ci'}}$ & ${\displaystyle \sum_{N\in\Z}\frac{e^{i\left(\theta-\theta'\right)N}}{2\pi}=\delta\left(\theta-\theta'\right)}$ & ${\displaystyle \int_{-\infty}^{\infty}\frac{dp}{2\pi}e^{i\left(x-x'\right)p}=\delta\left(x-x'\right)}$\tabularnewline
\bottomrule
\end{tabular}

\caption{\label{t:1}Comparison and summary of relations between conjugate
variables in $\protect\dfs$-, $\protect\fs$-, and $\protect\ft$-type
phase spaces. Most physically relevant phase spaces consist of combinations
of these three. An extension of the present analysis can be performed
for any (locally compact) Abelian group \cite{werner2016} and for
(compact) non-Abelian groups \cite{Lenz1990} (if one is willing to
allow commutators to be non-trivial operators, thereby dropping the
notion of a phase space). The (Abelian) groups involved here are $\protect\Z_{\protect\N}=\{-\left\lfloor \protect\N/2\right\rfloor ,\cdots,\left\lfloor (\protect\N-1)/2\right\rfloor \}$,
the circle group $\mathbb{T}=\{e^{i\theta}\,|\,\theta\in[-\pi,\pi[\}$,
the integers $\mathbb{Z}$, and the reals $\mathbb{R}$. The variables
$\protect\M\in\protect\Z$, $\vartheta\in[-\pi,\pi[$, and $\xi\in\mathbb{R}$.
In order to respect the domains of their respective conjugate variables,
the functionals $G$ and $K$ are respectively $\protect\N$- and
$2\pi$-periodic in their arguments. The canonical commutation relations
(CCR's) in quotes are incorrect due to not preserving domains, but
nevertheless give the right physical intuition. The label ``Bases''
is in quotes because $|\theta\protect\ket$, $|x\protect\ket$, and
$|p\protect\ket$ are not normalizable. See related tables in Refs.
\cite{Cotfas2012,werner2016}.}
\end{table*}

\subsection{Classical phase space}

\begin{figure}
\begin{centering}
\includegraphics[width=0.2\textwidth]{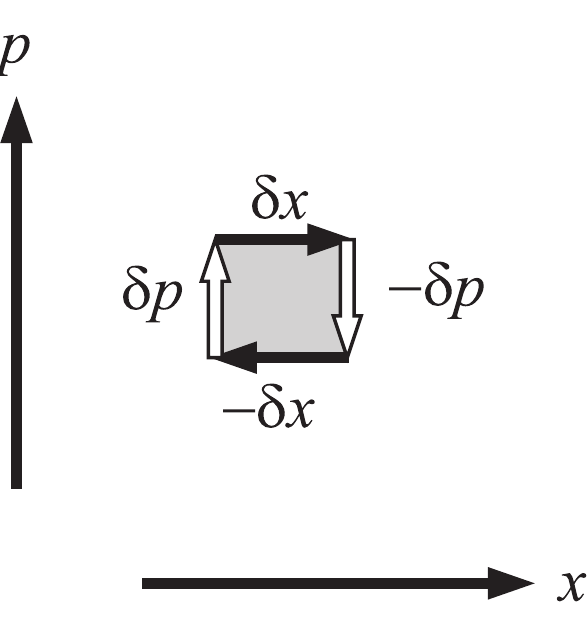}
\par\end{centering}
\caption{\label{fig:In-phase-space}In phase space, four elementary translations
along position and momentum define a closed circuit, here oriented
clockwise, and its corresponding enclosed area, represented here in
grey. Quantum physics associates a phase factor to an area in phase
space, measured in units of Planck's constant.}
\end{figure}

In classical physics, the phase space of a physical system with one
degree of freedom is a two-dimensional manifold spanned by two infinitesimal
translation generators, $T_{dx}$ and $T_{dp}$, acting on the conjugate
variables position and momentum, respectively. These translation operators
commute,
\begin{equation}
T_{dx}T_{dp}T_{dx}^{-1}T_{dp}^{-1}=\mathrm{Id}\,.\label{eq:trans}
\end{equation}
Starting with the system located at an origin point, it is possible
to reach a unique state in which the system is located at a well-defined
phase space point $\left(x,p\right)$ using a sequence of elementary
translations. The infinitesimal circuit associated with the sequence
of translations from Eq. (\ref{eq:trans}) defines a surface element
of phase space (see Fig. \ref{fig:In-phase-space}). Any observable
over phase space $f\left(x,p\right)$ evolves in time according to
Hamilton's equation, written here as 
\begin{equation}
df=\left\{ H,f\right\} dt\,,
\end{equation}
where the Poisson bracket of two functions $A\left(x,p\right)$ and
$B\left(x,p\right)$ of phase space is given by the exterior product
\begin{equation}
\left\{ A,B\right\} =\frac{\partial A}{\partial x}\frac{\partial B}{\partial p}-\frac{\partial A}{\partial p}\frac{\partial B}{\partial x}
\end{equation}
and $H\left(x,p\right)$ is the Hamiltonian function characterizing
the dynamics of system. Note that the Hamiltonian does not set the
nature of the degree of freedom itself, which is set by the topology
of phase space. For instance, the phase space of a 1-D massive particle
evolving in an $x^{2}$ potential (ideal harmonic oscillator) is a
flat plane whereas the phase space of a rigid pendulum or a rotor
is an infinite cylinder. While various phase spaces are used in classical
mechanics (depending on constraints one puts on a particle's motion),
there are only a few canonical topologies in quantum mechanics that
meaningfully extend the classical notion of conjugate variables.

\begin{figure}
\begin{centering}
\includegraphics[width=0.55\textwidth]{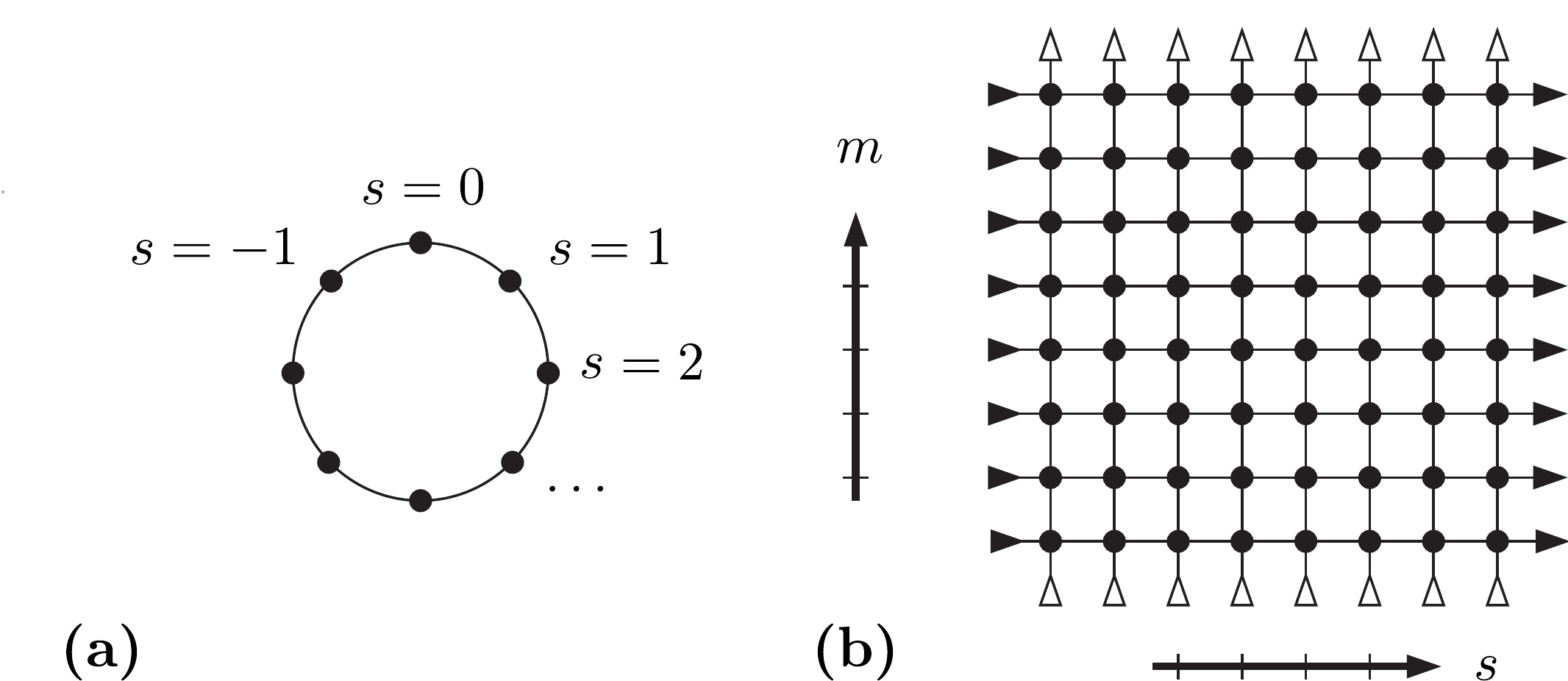}
\par\end{centering}
\caption{\label{fig:ring}\textbf{(a)} The $\protect\dfs$ phase space can
be thought of as consisting of a degree of freedom hopping from site
to site. The set of sites forms a ring. The position variable $s$
is a site index and is thus an integer modulo $\mathcal{N}$, the
number of sites along the ring. \textbf{(b)} If hopping between two
sites takes the universal same amount of time, phase space is fully
discrete, with the momentum $m$ belonging also to the set of integers
modulo $\mathcal{N}$. Both position and momentum have periodic boundary
conditions, as indicated by the set of black and white arrows, and
thus phase space has the topology of a torus.}
\end{figure}

In order to be sufficiently general in our quantum mechanical treatment,
instead of considering from the beginning a continuous phase space
with infinitesimal generators $T_{dx}$ and $T_{dp}$, we are going
to introduce \textit{finite} translation operators $T_{\delta x}$
and $T_{\delta p}$ and consider a topology of phase space where
\begin{eqnarray}
\left(T_{\delta x}\right)^{\N} & = & \left(T_{\delta p}\right)^{\N}=\mathrm{Id}\,,\label{eq:maxpow}
\end{eqnarray}
where $\N$ is a positive integer. Eventually, we will take the limits
$\delta x\rightarrow0$ , $\delta p\rightarrow0$ and $\mathcal{\N\rightarrow\infty}$
in varying ways. We thus consider that our 1-D degree of freedom,
instead of evolving continuously, hops from site to site, the set
of sites forming a ring graph shown in Fig. \ref{fig:ring}(a). The
position variable $s$ denotes the site index and is thus an integer
modulo $\N$, the total number of sites along the ring. If hopping
between two sites takes the same universal amount of time, phase space
is fully discrete, and the momentum $m$ also belongs to the set of
integers modulo $\N$. Because of periodic boundary conditions for
both position and momentum, as indicated by the set of black and white
arrows in Fig. \ref{fig:ring}(b), phase space has the topology of
a torus.

\subsection{Quantum phase space}

In quantum physics, two conjugate translation generators do not commute,
and one can categorize all possible relations into three cases:
\begin{equation}
T_{\delta x}T_{\delta p}T_{\delta x}^{-1}T_{\delta p}^{-1}\rightarrow C_{\delta x,\delta p}\coloneqq e^{-i\frac{\delta x\p}{\hbar}}e^{i\frac{\delta p\x}{\hbar}}e^{i\frac{\delta x\p}{\hbar}}e^{-i\frac{\delta p\x}{\hbar}}=\left\{ \begin{array}{c}
1\\
e^{-i\frac{\delta x\delta p}{\hbar}}\\
\text{operator not-commuting with translations}
\end{array}\right.\,.\label{non-commutation}
\end{equation}
The first case is obviously classical phase space. The second corresponds
to a \textit{quantum phase space} associated with a pair of Abelian
groups of conjugate translations, typical of oscillators and rotors.
The third case corresponds to spaces associated with non-Abelian groups,
such as the $SU(2)$ group associated with a spin $S$ (we discuss
later how the $S=\nicefrac{1}{2}$ representation does produce the
second case for particular $\d x,\d p$). In the following, we focus
on the second case, that of a circuit in phase space producing a phase
factor proportional to the area enclosed by the circuit, considering
different limits for the step sizes $\delta x$ and $\delta p$ and
the period $\N$. While we adhere to the definition of \textit{quantum}
\textit{phase space} which corresponds only to the second case \cite{werner2016},
we note that some of the properties we mention can also be extended
to the third case \cite{Lenz1990}.

Let us explore in detail the second case. Using Eqs. (\ref{eq:maxpow}-\ref{non-commutation})
to simplify $T_{\delta x}^{\N}T_{\delta p}^{\N}T_{\delta x}^{-\N}T_{\delta p}^{-\N}$,
we can see that
\begin{equation}
\left(C_{\delta x,\delta p}\right)^{\N}=\mathrm{Id}\,.
\end{equation}
Since $C_{\delta x,\delta p}$ commutes with every operator in the
algebra associated with the translations, we can represent it as
\begin{equation}
C_{\delta x,\delta p}\rightarrow e^{-i\frac{2\pi}{\N}}\,.\label{elementary-phase-shift}
\end{equation}
In our limit-taking procedures, the parameters $\delta x$, $\delta p$
and $\N$ do not vary independently, and we impose
\begin{equation}
\frac{2\pi}{\N}=\frac{\d x\d p}{\hbar}\,.\label{eq:eps}
\end{equation}
In other words, $C_{\delta x,\delta p}\rightarrow e^{-i\frac{\delta x\delta p}{\hbar}}$,
which corresponds to an elementary circuit in phase space accumulating
a phase shift given by the encircled area divided by Planck's constant,
in similarity with Bohr's old trajectory quantification rule.

Representing finite translations by exponentiation of translation
generators (which we denote in bold),\begin{subequations}
\begin{eqnarray}
T_{\delta x} & \rightarrow & e^{+i\frac{\delta x\p}{\hbar}}\\
T_{\delta p} & \rightarrow & e^{-i\frac{\delta p\x}{\hbar}}\,,
\end{eqnarray}
\end{subequations}we can rewrite the \textit{Weyl relation} as
\begin{equation}
e^{i\frac{\delta p\x}{\hbar}}e^{i\frac{\delta x\p}{\hbar}}e^{-i\frac{\delta p\x}{\hbar}}=e^{i\frac{\delta x}{\hbar}\left(\p-\d p\right)}=e^{-i\frac{\delta x\delta p}{\hbar}}e^{i\frac{\delta x\p}{\hbar}}\,.
\end{equation}
By naive expansion, we can recover the well-known result that the
Weyl relation for the $T^{\prime}s$ is equivalent to the \textit{canonical
commutation relation }(CCR) for the corresponding generating operators
\begin{equation}
\left[\x,\p\right]=i\hbar\,.\label{eq:comm}
\end{equation}
The generators $\x$ and $\p$ and the commutation relation above
define the algebra $\mathfrak{h}_{4}$ \cite{gilmore} (sometimes
referred to as the Heisenberg-Weyl algebra \cite{Klein1991}), which
has only infinite-dimensional representations. We will review how
the generators of motion for all common quantum-mechanical phase spaces,
some of which are finite-dimensional, nevertheless emulate $\mathfrak{h}_{4}$
(see Table \ref{t:1}, fourth row).

\subsection{Toroidal doubly-discrete quantum phase space ($\protect\dfs$)\label{sec:1:1}}

In the algebra representing the translation operators, the relations
(\ref{non-commutation}) and (\ref{elementary-phase-shift}) have
important consequences. Let us introduce the projectors $\Pi_{s}$
and $\Pi_{m}$ over respective regions of phase space $s\cdot\delta x$
and $m\cdot\delta p$, where $\d x$ and $\d p$ are some small intervals
which can be thought of as discretizations of ordinary continuous
phase space and
\begin{equation}
s,m\in\left\{ -\left\lfloor \N/2\right\rfloor ,\cdots,\left\lfloor (\N-1)/2\right\rfloor \right\} \eqqcolon\Z_{\N}\,.\label{eq:to-perform-this}
\end{equation}
We later vary the intervals $\delta x$ and $\delta p$ in a way which
converts the $\dfs$ phase space (and its associated Hilbert space)
into the $\fs$ and $\ft$ phase spaces. We have made the ranges of
$s,m$ ``two-sided'', i.e., defined them such that both their maximum
and minimum values are functions of $\N$, in order to properly perform
said procedures. The projectors (by definition, $\Pi_{s}=\Pi_{s}^{2}$
and $\Pi_{m}=\Pi_{m}^{2}$) satisfy\begin{subequations}
\begin{eqnarray}
T_{\delta x}\Pi_{s}T_{\delta x}^{-1} & = & \Pi_{s+1\text{mod}\N}\\
T_{\delta p}\Pi_{s}T_{\delta p}^{-1} & = & \Pi_{s}\\
T_{\delta p}\Pi_{m}T_{\delta p}^{-1} & = & \Pi_{m+1\text{mod}\N}\\
T_{\delta x}\Pi_{m}T_{\delta x}^{-1} & = & \Pi_{m}\,.
\end{eqnarray}
\end{subequations}Quantum-mechanically, due to the non-commutation
of conjugate translation operators, the product of projectors on position
and momentum $\Pi_{s}\Pi_{m}$ can no longer be a projector (as in
the classical case). Thus, $(\Pi_{s}\Pi_{m})^{2}\neq\Pi_{s}\Pi_{m}$
since the projectors do not commute and it is not possible to identify
a phase space point state with both definite position $s$ and momentum
$m$. As a consequence of projections on $m$ conflicting with projections
on $s$, we have the discrete Fourier relations
\begin{equation}
\left\vert m\right\rangle =\frac{1}{\sqrt{\N}}\sum_{\ci\in\Z_{\N}}e^{i\frac{2\pi sm}{\N}}\left\vert s\right\rangle \,\,\,\,\,\,\,\,\,\,\,\,\,\,\,\,\,\,\text{and}\,\,\,\,\,\,\,\,\,\,\,\,\,\,\,\,\,\,\left\vert s\right\rangle =\frac{1}{\sqrt{\N}}\sum_{\ri\in\Z_{\N}}e^{-i\frac{2\pi ms}{\N}}\left\vert m\right\rangle \,,\label{eq:defrho}
\end{equation}
where the position state vectors $\left\vert s\right\rangle $ and
momentum state vectors $\left\vert m\right\rangle $ are defined as
the eigenvectors of the projectors with eigenvalues 0 and 1: 
\begin{equation}
\Pi_{s}\left\vert s^{\pr}\right\rangle =\delta_{s^{\pr},s}\left\vert s^{\pr}\right\rangle \,\,\,\,\,\,\,\,\,\,\,\,\,\,\,\,\,\,\text{and}\,\,\,\,\,\,\,\,\,\,\,\,\,\,\,\,\,\,\Pi_{m}\left\vert m^{\prime}\right\rangle =\delta_{m^{\prime},m}\left\vert m^{\prime}\right\rangle \,.
\end{equation}
The projectors in the position basis and in the momentum basis both
resolve the identity,
\begin{eqnarray}
\sum_{\ci\in\Z_{\N}}\left\vert s\right\rangle \left\langle s\right\vert  & = & \sum_{\ci\in\Z_{\N}}\Pi_{s}=\id\,\,\,\,\,\,\,\,\,\,\,\,\,\,\,\text{and}\,\,\,\,\,\,\,\,\,\,\,\,\,\,\,\sum_{\ri\in\Z_{\N}}\left\vert m\right\rangle \left\langle m\right\vert =\sum_{\ri\in\Z_{\N}}\Pi_{m}=\id\,,
\end{eqnarray}
and the overlap amplitude of the basis vectors is a constant,\footnote{A pair of bases for which the norm of the overlap between their constituents
is independent of the basis labels is called \textit{mutually unbiased}
\cite{DURT2010}.}
\begin{equation}
\left\langle s|m\right\rangle =\frac{1}{\sqrt{\N}}e^{i\frac{2\pi sm}{\N}}\,\,\,\,\,\,\,\,\,\,\,\,\,\,\text{implies}\,\,\,\,\,\,\,\,\,\,\,\,\,\,\left\vert \left\langle s|m\right\rangle \right\vert ^{2}=\frac{1}{\N}\,.
\end{equation}
There are thus only $\N-1$ independent projectors whereas there are
$\N^{2}$ orthogonal observables (counting the identity as the constant
observable). Classically, there would be also $\N^{2}$ orthogonal
observables, but there would be also as much as $\N^{2}-1$ projectors.
Thus, in quantum mechanics, the number of independent pure states
is much less than the number of properties that can be acquired from
them!

Let us now introduce the conjugate variables that label states of
fixed position and momentum,
\begin{equation}
\c=\sum_{\ci\in\Z_{\N}}s\left\vert s\right\rangle \left\langle s\right\vert \,\,\,\,\,\,\,\,\,\,\,\,\,\,\,\text{and}\,\,\,\,\,\,\,\,\,\,\,\,\,\,\,\r=\sum_{\ri\in\Z_{\N}}m\left\vert m\right\rangle \left\langle m\right\vert \,,\label{eq:dfsconj}
\end{equation}
and quantify their conjugate nature. These operators label the columns
and the rows of Fig. \ref{fig:ring}b, but again the non-commutation
of these operators prevents simultaneously assigning a fixed position
and momentum to the points in the corresponding classical space. The
quantum or discrete Fourier transform operator going from the momentum
basis to the position basis is
\begin{equation}
\F_{\dfs}\coloneqq\sum_{\ci\in\Z_{\N}}\left\vert s\right\rangle \left\langle m=s\right\vert =\frac{1}{\sqrt{\N}}\sum_{\ci\in\Z_{\N}}\sum_{\ci^{\prime}\in\Z_{\N}}e^{i\frac{2\pi ss^{\prime}}{\N}}|s\ket\bra s^{\prime}|\,.\label{eq:dfsfourier}
\end{equation}
The position and momentum operators are related by this transform
via
\begin{equation}
\F_{\dfs}^{\dg}\,\r\,\F_{\dfs}=\c\,\,\,\,\,\,\,\,\,\,\,\,\,\,\,\,\text{and}\,\,\,\,\,\,\,\,\,\,\,\,\,\,\,\,\F_{\dfs}^{\dg}\,\c\,\F_{\dfs}=-\r\,.
\end{equation}
Performing the discrete Fourier transform twice yields the $\dfs$
parity operator 
\begin{equation}
\F_{\dfs}^{2}=\sum_{s\in\Z_{\N}}\left|-s\right\rangle \bra s|\,.\label{eq:parity}
\end{equation}
This important operator takes $\c,\r$ to $-\c,-\r$ modulo $\N$. 

We can readily make contact with standard Fourier analysis by linking
a quantum pure state $|\psi\ket$ to a function of a discrete periodic
variable $\psi_{m}$. Namely, writing $|\psi\ket$ using the position
basis and looking at the overlap of $|\psi\ket$ with a momentum eigenstate
$|m\ket$ produces the discrete Fourier series of $\psi_{m}$:
\begin{equation}
|\psi\ket=\sum_{\ci\in\Z_{\N}}\psi_{s}|s\ket\,\,\,\,\,\,\,\,\,\Leftrightarrow\,\,\,\,\,\,\,\,\,\psi_{m}\coloneqq\bra m|\psi\ket=\frac{1}{\sqrt{\N}}\sum_{\ci\in\Z_{\N}}\psi_{s}e^{-i\frac{2\pi sm}{\N}}\,.\label{eq:fourier}
\end{equation}
The same holds for the dual momentum basis.

In the position state vector basis, translations in position and in
momentum are explicitly given by\footnote{These were introduced first by Sylvester in the 19th century \cite{sylvester}
and applied to quantum mechanics by von Neumann \cite{vonneumann},
Weyl \cite{weyl_book}, and Schwinger \cite{schwinger}. They have
been called Schwinger bases, Weyl operators \cite{werner2016}, Pauli
operators \cite{gottesman2001}, generalized spin \cite{gsm} or Pauli
\cite{Kibler2009} matrices, and 't Hooft generators or clock-and-shift
matrices \cite{Sachse2006}.}\begin{subequations}
\begin{eqnarray}
T_{\delta x} & \rightarrow & \X\coloneqq e^{-i\frac{2\pi}{\N}\r}=\sum_{\ci\in\Z_{\N}}\left\vert s+1\text{mod}\N\right\rangle \left\langle s\right\vert \label{eq:X}\\
T_{\delta p} & \rightarrow & \ZZ\coloneqq e^{+i\frac{2\pi}{\N}\c}=\sum_{\ci\in\Z_{\N}}e^{+i\frac{2\pi s}{\N}}\left\vert s\right\rangle \left\langle s\right\vert \,,\label{eq:Z}
\end{eqnarray}
\end{subequations}and such operators naturally perform displacements
along the ring of sites in either position or momentum cross-sections
of phase space:\footnote{The set $\{e^{i\frac{2\pi j}{\N}}e^{+i\frac{2\pi k}{\N}\c}e^{-i\frac{2\pi l}{\N}\r}\}_{j,k,l\in\Z_{\N}}$
forms a group, called the Generalized Pauli Group \cite{Kibler2009},
and an algebra, sometimes called the non-commutative torus \cite{Landi2001}.}
\begin{eqnarray}
\X^{\dg}\c\X & = & \left(\c+1\right)\text{mod}\N\,\,\,\,\,\,\,\,\,\,\,\,\,\,\,\,\,\text{and}\,\,\,\,\,\,\,\,\,\,\,\,\,\,\,\,\,\ZZ^{\dg}\r\ZZ=\left(\r+1\right)\text{mod}\N\,.\label{eq:shifts}
\end{eqnarray}
Strictly speaking, we cannot relate the Weyl relation
\begin{equation}
\X\ZZ\X^{\dg}\ZZ^{\dg}=e^{-i\frac{2\pi}{\N}}\label{eq:grppauli}
\end{equation}
to a CCR since expanding $\X,\ZZ$ to first order will violate our
imposed domains on $\c,\r$ \cite{Hall2013}. In other words, $\c,\r$
can only be inside functions that are periodic in $\N$. However,
we can depart from mathematical rigor and represent the above Weyl
relation as the CCR
\begin{equation}
\text{\textquotedblleft}\left[\c,\r\right]=i\frac{\N}{2\pi}\text{\textquotedblright}\,,\label{eq:dfscr}
\end{equation}
which is analogous to the continuous case (\ref{eq:comm}). Moreover,
if we express everything in terms of
\begin{equation}
\x_{\dfs}\coloneqq\d x\,\c\,\,\,\,\,\,\,\,\,\,\,\,\,\,\,\,\,\text{and}\,\,\,\,\,\,\,\,\,\,\,\,\,\,\,\,\,\p_{\dfs}\coloneqq\d p\,\r\label{eq:red}
\end{equation}
and use our imposed phase space area constraint (\ref{eq:eps}), the
translations can be expressed as $\X=e^{-i\frac{\delta x}{\mathcal{\hbar}}\p_{\dfs}}$
and $\ZZ=e^{+i\frac{\delta p}{\mathcal{\hbar}}\x_{\dfs}}$ and Eq.
(\ref{eq:dfscr}) recovers the ordinary CCR $\text{\textquotedblleft}[\x_{\dfs},\p_{\dfs}]=i\hbar\text{\textquotedblright}$
(\ref{eq:comm}).

We briefly describe a Wigner function representation for $\dfs$ for
$\N$ being odd \cite{Vourdas2017}. Recall that $\ft$ Wigner functions
can be expressed in terms of the trace of a density matrix with a
certain displaced parity operator; we provide the expression later
in Eq. (\ref{eq:wigft}). In $\dfs$, an analogous expression is
\begin{equation}
W_{\dfs}\left(S,M\right)=\frac{1}{\N}\sum_{s=0}^{\N-1}\bra s|\rho\D_{\dfs}^{S,M}\F_{\dfs}^{2}\D_{\dfs}^{S,M\dg}|s\ket=\frac{1}{\N}\tr\left\{ \rho\D_{\dfs}^{S,M}\F_{\dfs}^{2}\D_{\dfs}^{S,M\dg}\right\} \,,\label{eq:wigtor}
\end{equation}
where $S,M\in\Z_{\N}$, $\rho$ is an $\N\times\N$ density matrix,
$\F_{\dfs}^{2}$ (\ref{eq:parity}) is the $\dfs$ parity operator,
and $\D_{\dfs}^{S,M}\coloneqq e^{-i\frac{\pi SM}{\N}}e^{+i\frac{2\pi M}{\N}\c}e^{-i\frac{2\pi S}{\N}\r}$
is the $\dfs$ displacement operator. The Wigner function conveniently
takes real values over phase space and thus shares some of the properties
of classical probability distributions, despite not always having
positive values.

Properties of this $\dfs$ fully discrete phase space for general
$\N$ are summarized in the first column of Table \ref{t:1}. We have
only introduced the bare-bones framework, and there are many more
quantities that can be defined in $\dfs$, including coherent states
\cite{Galetti1996,Ruzzi2005}, squeezed states \cite{Marchiolli2007},
and quantum codes (\cite{Ketkar2006}; \cite{gottesman2001}, Sec.
II). There are also plenty of other ways to visualize states \cite{Galetti1988,Ruzzi2000,Gibbons2004,Marchiolli2005,Ferrie2011,Marchiolli2012,Tilma2016,Ligabo2016}.
We refer the reader to Refs. \cite{imoto_book,DelaTorre2003,Bengtsson2017,Vourdas2017}
for further introductory reading.

\subsubsection*{The $\protect\N=2$ case \textemdash{} a spin one-half system}

It is important to realize that the case $\N=2$ is that of the ubiquitous
spin-$\nicefrac{1}{2}$. In that case,\begin{subequations}
\begin{eqnarray}
\left(T_{\delta x}\right)^{2} & = & \id\\
\left(T_{\delta p}\right)^{2} & = & \id\\
T_{\delta x}T_{\delta p}T_{\delta x}^{-1}T_{\delta p}^{-1} & = & T_{\delta x}T_{\delta p}T_{\delta x}T_{\delta p}=e^{-i\pi}\id=-\id\,.
\end{eqnarray}
\end{subequations}Therefore, $T_{\delta x}=\sigma_{x}$ and $T_{\delta p}=\sigma_{z}$.
Since $\sigma_{x}=e^{i\frac{\pi}{2}\left(\sigma_{x}-\id\right)}$
and same for $\s_{z}$, we have (modulo 2)
\[
\c=\begin{pmatrix}-1 & 0\\
0 & 0
\end{pmatrix}=\frac{\sigma_{z}-\id}{2}\,\,\,\,\,\,\,\,\,\,\,\,\,\,\,\,\,\text{and}\,\,\,\,\,\,\,\,\,\,\,\,\,\,\,\,\,\r=\frac{\sigma_{x}-\id}{2}\,.
\]
The Fourier transform corresponds to the well-known Hadamard transform
and the parity operator (\ref{eq:parity}) $\F_{\dfs}^{2}=\id$ is
trivial. We thus see that at the particular angle $\theta=\nicefrac{\pi}{2}$,
the expression $e^{i\theta\sigma_{x}}e^{i\theta\sigma_{z}}e^{-i\theta\sigma_{x}}e^{-i\theta\sigma_{z}}$
does produce a constant not equal to one \textemdash{} the second
case in Eq.~(\ref{non-commutation}). This does not occur for any
other values of $\theta$. Recalling that the generators $\{S_{x},S_{y},S_{z}\}$
of the $\mathfrak{su}(2)$ Lie algebra can be realized in a space
of dimension $2S+1$ given a spin $S$ \cite{schiff}, the $S=\nicefrac{1}{2}$
case at $\theta=\nicefrac{\pi}{2}$ is the only time that spin rotations
$\{e^{i\theta S_{x}},e^{i\theta S_{z}}\}$ and $\dfs$ translation
operators $\{\X,\ZZ\}$ coincide. Therefore, procedures involving
representations of $\mathfrak{su}(2)$ for $S>\nicefrac{1}{2}$, such
as spin-coherent states \cite{puri}, the Holstein-Primakoff transformation,
and its associated Lie-algebraic contraction ($\mathfrak{u}\left(2\right)\rightarrow\mathfrak{h}_{4}$;
see, e.g., \cite{Arecchi1972,ATAKISHIYEV2003,gilmore}), are \textit{not}
directly connected to the phase space analysis discussed here for
$\N>2$.

\subsection{Cylindrical singly-discrete quantum phase space ($\protect\fs$)\label{subsec:Cylindrical-singly-discrete-quan}}

We now take the limit $\N\rightarrow\infty$, first considering the
case where 
\begin{equation}
\delta x=\frac{2\pi}{\N}\rightarrow0\,\,\,\,\,\,\,\,\,\,\,\,\,\,\,\,\,\,\,\,\,\text{and}\,\,\,\,\,\,\,\,\,\,\,\,\,\,\,\,\,\,\,\,\,\d p=\con\,,
\end{equation}
invoking a universal constant $\mathcal{C}$. Then we can introduce
conjugate variables
\begin{equation}
\k\leftarrow\frac{2\pi}{\N}\,\c\,\,\,\,\,\,\,\,\,\,\,\,\,\,\,\,\,\,\,\,\,\text{and}\,\,\,\,\,\,\,\,\,\,\,\,\,\,\,\,\,\,\,\,\,\NN\leftarrow\r\,.
\end{equation}
The operators $\k$ and $\NN$ take their eigenvalues in the set of
angles (compact set of reals modulo $2\pi$) and the set of all integers,
respectively:
\begin{eqnarray}
\theta & \in & [-\pi,\pi[\,\,\,\,\,\,\,\,\,\,\,\,\,\,\,\,\text{and}\,\,\,\,\,\,\,\,\,\,\,\,\,\,\,\,N\in\mathbb{Z}\,,
\end{eqnarray}
hence the renaming of $\r$ into $\NN$. 

In terms of wavefunctions, the limit as the number of points in the
discretization $\N\rightarrow\infty$ is equivalent to the standard
limit in which the discrete Fourier series of a \textit{discrete}
periodic wavefunction $\psi_{m}$ is transformed into the ordinary
Fourier series of a \textit{continuous} periodic function (see \cite{kbwolf},
Sec. 3.4.5). In terms of the new conjugate variables, Eq. (\ref{eq:fourier})
becomes
\begin{equation}
\!\!\!\!\!\!\!\!\!\!\psi_{m}=\frac{1}{\sqrt{\N}}\sum_{s}\psi_{s}e^{-i\frac{2\pi sm}{\N}}=\frac{1}{\sqrt{2\pi}}\sum_{\theta}\psi(\theta)e^{-i\theta N}\frac{2\pi}{\N}\,\,\,\xrightarrow{\N\rightarrow\infty}\,\,\,\frac{1}{\sqrt{2\pi}}\int_{-\pi}^{\pi}\psi(\theta)e^{-i\theta N}d\theta\,,\label{eq:fsbasis}
\end{equation}
where we have rescaled the coefficients as $\psi(\theta)\coloneqq\psi_{m}\sqrt{\nicefrac{\N}{2\pi}}$.
In the large $\N$ limit, this position-basis expansion of $\psi$
becomes an integral over the angle $\theta$. In this \textit{infinite
ladder} or $\fs1$ limit ($\c,\r\rightarrow\k,\NN$), the circle labeled
by eigenvalues of $\r$ is essentially ``cut open'' and turns into
the unbounded integer-valued variable $\NN$, while at the same time
$\c$ is absorbed into the dense and bounded variable $\k$ conjugate
to $\NN$. Of course, one could have instead done $\c,\r\rightarrow\NN,\k$,
which we call the the \textit{infinitely dense circle} or $\fs2$\textit{
}limit. How the remaining properties of $\dfs$ transform in this
limit are listed in Table \ref{t:1}. The concepts discussed for $\dfs$
in the text, such as coherent states \cite{Kowalski1996} and Wigner
functions \cite{Berry1977,Mukunda1979}, also naturally carry over
to $\fs$ (see also, e.g., \cite{Zhang2003,Ruzzi2006}).

In the $\fs$1 limit, $\dfs$ position and momentum eigenstates (\ref{eq:defrho})
become
\begin{eqnarray}
\left\vert \theta\right\rangle  & = & \frac{1}{\sqrt{2\pi}}\sum_{N\in\mathbb{Z}}e^{-i\theta N}\left\vert N\right\rangle \,\,\,\,\,\,\,\,\,\,\,\,\,\,\,\,\,\,\text{and}\,\,\,\,\,\,\,\,\,\,\,\,\,\,\,\,\,\,\left\vert N\right\rangle =\frac{1}{\sqrt{2\pi}}\int_{-\pi}^{\pi}d\theta\,e^{i\theta N}\left\vert \theta\right\rangle \,,\label{eq:theta}
\end{eqnarray}
respectively. Here, we encounter states which are normalizable only
in the ``Dirac'' or ``continuous'' sense as well as the technicality
that the orthonormality relation has to be $2\pi$-periodic:
\begin{equation}
\bra\theta|\theta^{\pr}\ket=\frac{1}{2\pi}\sum_{N\in\Z}e^{-i\left(\theta^{\prime}-\theta\right)N}=\sum_{N\in\Z}\d\left(\theta^{\prime}-\theta-2\pi N\right)\eqqcolon\d^{\left(2\pi\right)}\left(\theta^{\pr}-\theta\right)\,.
\end{equation}
Above, we define the $2\pi$-periodic $\d$-function in order to make
sure that we can use any values of $\theta$ \cite{Raynal2010}. However,
if we restrict ourselves to using only $\theta\in[-\pi,\pi[$, as
in Table \ref{t:1}, the $\d^{\left(2\pi\right)}$-function reduces
to the ordinary $\d$-function. Since $|\theta\ket$ are not normalizable,
they technically do not belong to the function space $L^{2}(-\pi,\pi)$
associated with $\fs$, i.e., the space of functions $f(\theta)$
such that $\int d\theta|f(\theta)|^{2}<\infty$ (\cite{Hall2013},
Sec. 6.6).

Another consequence of domains and similar to the $\dfs$ case, the
Weyl relation between translations in $\NN$ and $\k$ does not imply
a proper CCR (see Ref. \cite{Hall2013}, Sec. 12.2). Assuming that
restrict ourselves to using only $\theta\in[-\pi,\pi[$, functions
of $\k$ must be $2\pi$-periodic in order to preserve its domain.
Therefore, $\k$ and its powers cannot act on states alone. If we
ignore this fact and calculate the variances of states $|n\ket$ in
$\k$ and $\NN$, then we will see that the former yields a finite
number while the latter is zero. This violates Heisenberg's uncertainty
relation and thus the conventional CCR (we list what the CCR would
have been if we did not worry about domains in Table \ref{t:1}).

Application for the $\fs$ phase space include i) the quantum rotor
\cite{Raynal2010,wenbook}, where $N$ ($\theta$) labels the angular
momentum (position) of the rotor, ii) the motion of an electronic
excitation in the periodic potential of crystal, where $N$ is the
site index, assuming the crystal to be infinite, which makes $\theta$
analogous to the pseudo momentum in band-theory \cite{bandtheory},
or iii) the dynamics of a Josephson junction between two isolated
islands, like in the Cooper pair box \cite{girvinbook,Devoret1995},
where $\theta$ is the phase difference between the two superconductors
on either side of the junction and $N$ the number of Cooper pairs
having tunneled across the junction. 

\subsection{Flat-plane fully continuous quantum phase space ($\protect\ft$)\label{subsec:Flat-plane-fully}}

We again take the limit $\N\rightarrow\infty$, but now consider the
case where both
\begin{equation}
\d x=\d p=\sqrt{\frac{2\pi}{\N}}
\end{equation}
approach zero. Thus, while the whole of phase space has a number of
points growing as $\N^{2}$, an area of order $\hbar$ will harbor
of order $\N$ points and can still be considered continuous. Note
that we did not have to split $\nicefrac{2\pi}{\N}$ into two identical
factors; any splitting $\d x=(\nicefrac{2\pi}{\N})^{1-\frac{\epsilon}{2}}$
and $\d p=(\nicefrac{2\pi}{\N})^{\frac{\epsilon}{2}}$ for $0<\epsilon<2$
is sufficient \cite{Ruzzi2002a,Ruzzi2002}. Keeping with an even splitting,
we introduce new conjugate variables
\begin{equation}
\x\leftarrow\sqrt{\frac{2\pi}{\N}}\,\c\,\,\,\,\,\,\,\,\,\,\,\,\,\,\,\,\,\,\,\,\,\,\,\,\,\,\,\,\,\text{and}\,\,\,\,\,\,\,\,\,\,\,\,\,\,\,\,\,\,\,\,\,\,\,\,\,\,\,\,\,\p\leftarrow\sqrt{\frac{2\pi}{\N}}\,\r\,,\label{eq:cv}
\end{equation}
which become ordinary position and momentum in the large $\N$ limit.
We had already seen from Eq. (\ref{eq:red}) that this type of redefinition
recovers the original commutation relation $[\x,\p]=i\hbar$ (\ref{eq:comm}).
In terms of wavefunctions, this is equivalent to the standard limit
in which the discrete Fourier series of a periodic wavefunction $\psi_{m}$
is transformed into the continuum Fourier series as the functions
period $\N\rightarrow\infty$ (see \cite{kbwolf}, Sec. 3.4.5). In
terms of the new conjugate variables, Eq. (\ref{eq:fourier}) becomes
\begin{equation}
\!\!\!\!\!\!\!\!\!\!\psi_{m}=\frac{1}{\sqrt{\N}}\sum_{s}\psi_{s}e^{-i\frac{2\pi sm}{\N}}=\frac{1}{\sqrt{2\pi}}\sum_{p}\psi(p)e^{-ixp}\d p\,\,\,\xrightarrow{\N\rightarrow\infty}\,\,\,\frac{1}{\sqrt{2\pi}}{\displaystyle \int_{-\infty}^{\infty}}\psi(p)e^{-ixp}dp\,,\label{eq:sum}
\end{equation}
where $\psi(p)\coloneqq\psi_{s}$. Since $\d p\rightarrow0$ for large
$\N$, the above sum over $p$ (\ref{eq:sum}) becomes an integral
over $\mathbb{R}$. This completes the limit-taking procedure $\c,\r\rightarrow\x,\p$.
The properties of this continuous flat phase space are summarized
in the last column of Table \ref{t:1}. Just like $\F_{\dfs}$ (\ref{eq:dfsfourier}),
we can write the Fourier transform $\F_{\ft}$ as a standalone operator:
\begin{equation}
\F_{\ft}=\int_{-\infty}^{\infty}dx\left\vert x\right\rangle \left\langle p=x\right\vert =\frac{1}{\sqrt{2\pi}}\int_{-\infty}^{\infty}dx\int_{-\infty}^{\infty}dx^{\prime}e^{ixx^{\prime}}|x\ket\bra x^{\prime}|\,.\label{eq:ftfourier}
\end{equation}
One can easily confirm that $\F_{\ft}^{2}$ is the parity operator
taking $\x,\p\rightarrow-\x,-\p$. Note that eigenfunctions $|x\ket,|p\ket$
of position and momentum are, like $|\theta\ket$ (\ref{eq:theta}),
not normalizable and therefore not in the space of physical quantum
states $L^{2}(\mathbb{R})$ (\cite{Hall2013}, Sec. 6.6).

The $\ft$ Wigner function $W_{\ft}$ can then defined, analogous
to $W_{\dfs}$ (\ref{eq:wigtor}), in terms of a $\ft$ displacement
operator $\D_{\ft}^{X,P}\coloneqq e^{-iXP}e^{2iP\x}e^{-2iX\p}$ and
the parity operator $\F_{\ft}^{2}$. One can easily confirm that $\F_{\ft}^{2}$
takes $\x,\p\rightarrow-\x,-\p$. Letting $X,P\in\mathbb{R}$ and
following Appx. A.2.1 of Ref. \cite{catbook} yields 
\begin{equation}
W_{\ft}\left(X,P\right)=\frac{2}{\pi}\int_{-\infty}^{\infty}dx\bra x|\rho\D_{\ft}^{X,P}\F_{\ft}^{2}\D_{\ft}^{X,P\dg}|x\ket=\frac{2}{\pi}\tr\left\{ \rho\D_{\ft}^{X,P}\F_{\ft}^{2}\D_{\ft}^{X,P\dg}\right\} \,.\label{eq:wigft}
\end{equation}

Now that we have performed the $\dfs\rightarrow\fs$ and $\dfs\rightarrow\ft$
limit-taking procedures, all that is left to complete the connections
between them is the $\fs\rightarrow\ft$ limit. Recall that the $\fs$
variables are the angular $\k$ and integer $\NN$. To perform the
limit, we introduce a length scale $\L$ which rescales the periodicity
of $\k$ and take this scale to infinity. The new variables this time
are
\begin{equation}
\x\leftarrow\frac{\L}{2\pi}\k\,\,\,\,\,\,\,\,\,\,\,\,\,\,\,\,\,\,\,\,\,\,\,\,\,\,\,\,\,\text{and}\,\,\,\,\,\,\,\,\,\,\,\,\,\,\,\,\,\,\,\,\,\,\,\,\,\,\,\,\,\p\leftarrow\frac{2\pi}{\L}\NN\,.\label{eq:thirdlim}
\end{equation}
The first redefinition transforms the already continuous variable
$\k$ into an unbounded variable while the second transforms the already
unbounded variable $\NN$ into a continuous one (since its intervals
$\d p\coloneqq\frac{2\pi}{\L}$ go to zero). In terms of the new conjugate
variables, the $|\theta\ket$ component of $|\psi\ket$ expanded in
the $|N\ket$ basis becomes
\begin{equation}
\!\!\!\!\!\!\!\!\!\!\psi(\theta)\coloneqq\bra\theta|\psi\ket=\frac{1}{\sqrt{2\pi}}\sum_{N\in\Z}\psi_{N}e^{i\theta N}=\frac{1}{\sqrt{2\pi}}\sum_{p}\psi(p)e^{-ixp}\d p\,\,\,\xrightarrow{\L\rightarrow\infty}\,\,\,\frac{1}{\sqrt{2\pi}}{\displaystyle \int_{-\infty}^{\infty}}\psi(p)e^{-ixp}dp\,,
\end{equation}
where we define the rescaled coefficients $\psi(p)\coloneqq\frac{\L}{2\pi}\psi_{N}$.
This completes the last limit $\k,\NN\rightarrow\x,\p$, which is
based on the well-known conversion of a Fourier series of a periodic
function $\psi(\theta)$ into a Fourier transform by taking the function's
periodicity $\L$ to infinity.

Since the periodicity of both position and momentum goes to infinity,
these $\dfs\rightarrow\ft$ and $\fs\rightarrow\ft$ limit-taking
procedures are well adapted to studies of harmonic and weakly anharmonic
oscillators and to expansions of periodic functions of operators.
Letting $x_{ZPF}$ and $p_{ZPF}$ be standard deviations of the zero
point fluctuations of the oscillator, we can introduce the operators
$\aa$ and $\aa^{\dg}$ such that 
\begin{equation}
\x=x_{ZPF}\left(\aa+\aa^{\dg}\right)\,,\,\,\,\,\,\,\,\,\,\,\,\,\,\,\,\,\,\,\,\p=p_{ZPF}\left(\aa-\aa^{\dg}\right)/i\,,\,\,\,\,\,\,\,\,\,\,\,\,\,\,\,\,\,\,\,\text{and}\,\,\,\,\,\,\,\,\,\,\,\,\,\,\,\,\,\,\,\left[\aa,\aa^{\dg}\right]=\id\,.
\end{equation}
Then, the number of action quanta in the system is the operator
\begin{equation}
\ph=\aa^{\dg}\aa\,.
\end{equation}
In general, the Hamiltonian is not a simple function of $\ph$, but
remains a balanced function of $\aa$ and $\aa^{\dg}$. Using this
notation, the Fourier transform and parity operator are simply
\begin{equation}
\F_{\ft}=e^{i\frac{\pi}{2}\aa^{\dg}\aa}\,\,\,\,\,\,\,\,\,\,\,\,\,\,\,\,\,\,\,\,\,\text{and}\,\,\,\,\,\,\,\,\,\,\,\,\,\,\,\,\,\,\,\,\,\F_{\ft}^{2}=\left(-1\right)^{\aa^{\dg}\aa}\,.\label{eq:parity-1}
\end{equation}
Note that in contrast with the situation with the pair $\NN$ and
$\k$, there is no conjugate quantum operator for $\ph$ satisfying
all of the properties in Table \ref{t:1} (although there is an operator
satisfying some of the properties \cite{Garrison1970}). This is due
to the fact that the polar representation of even a flat plane is
singular when the radius is zero (equivalently, the eigenvalues of
$\ph$ are bounded from below). This effect also obstructs us from
creating an orthonormal basis of phase states for quantum optical
applications (see Ref. \cite{Lynch1995} or Ref. \cite{schleich},
Problem 8.4). 

\begin{table*}[t]
\begin{tabular}[t]{lccc}
\toprule 
 & $\dfs$ & $\fs$ & $\ft$\tabularnewline
\midrule 
Sec. \ref{sec:sho} & Harper equation \cite{azbel,Hofstadter1976} & ~~Almost Mathieu equation \cite{Avila2009}~~ & Harmonic oscillator\tabularnewline
Sec. \ref{sec:Baxter} & ~~Baxter $Z_{\N}$ parafermionic spin chain \cite{Baxter1989}~~ & Rotor Baxter chain & Coupled-oscillator chain\tabularnewline
Sec. \ref{sec:rabi2} & $\N$-state Rabi model \cite{rabi,rabi2,pub009} & Rotor-oscillator Rabi model & Optomechanical Hamiltonian\tabularnewline
Sec. \ref{sec:toric-code} & Kitaev $\Z_{\N}$ toric code \cite{Kitaev2003,Wen2003,Bullock2007} & Rotor toric code & ~~CV toric code \cite{Zhang2008}~~\tabularnewline
Sec. \ref{sec:Haah-cubic-code} & Haah $\Z_{\N}$ cubic code \cite{Haah2011} & Rotor Haah code & CV honeycomb model\tabularnewline
Sec. \ref{sec:Kitaev-model} & Kitaev honeycomb model \cite{Kitaev2006,Barkeshli2015,Fendley2012} & Rotor honeycomb model & CV Haah code\tabularnewline
\bottomrule
\end{tabular}

\caption{\label{t:2}The first column lists models to which we apply the limit-taking
procedures of Sec. \ref{sec:Continuum-and-rotor}, yielding the rotor
($\protect\fs$) and continuous-variable ($\protect\ft$) models in
the last two columns.}
\end{table*}

\section{Continuum and rotor limit-taking procedures\label{sec:Continuum-and-rotor}}

The above formulations of $\fs$ and $\ft$ from $\dfs$ are only
done on the level of the Hilbert space. When it comes to applying
them to Hamiltonians, there are some additional subtleties which have
to be dealt with. In an attempt to resolve such subtleties, let us
demonstrate our slightly generalized limit-taking procedures on a
general $\dfs$-type Hamiltonian. Since we saw that there are two
ways to take the $\dfs\rightarrow\fs$ limit in Sec. \ref{subsec:Cylindrical-singly-discrete-quan},
the limits we consider below are summarized in the following diagram:
\begin{equation}
\begin{array}{ccccc}
 &  & \fs1\\
 & \nearrow\!\!\!\!\!\!\! &  & \!\!\!\!\!\!\!\searrow\\
\dfs\!\!\!\!\!\!\! &  & \rightarrow &  & \!\!\!\!\!\!\!\ft\\
 & \searrow\!\!\!\!\!\!\! &  & \!\!\!\!\!\!\!\nearrow\\
 &  & \fs2
\end{array}\,.\label{eq:diagram}
\end{equation}
We start with a $\dfs$ Hamiltonian that can be written as
\begin{equation}
H_{\dfs}=-\frac{1}{2}\sum_{k}f_{k}\left(\frac{2\pi\M}{\N}\c\right)g_{k}\left(\frac{2\pi\L}{\N}\r\right)+H.c.\,,\label{eq:torgen}
\end{equation}
where $0<\M,\L<\N$ modulate the hopping length scales for the respective
variables and $f_{k},g_{k}$ are analytic $\N$-periodic functions
of $\c,\r$ (\ref{eq:dfsconj}). Simpler versions of the $\dfs\rightarrow\ft$
limit, which are applicable for all but one of the models we consider,
can be performed with $\M=\L=1$. However, these scales are necessary
to be able to obtain $\ft$ via $\fs$, so we keep them for now.

Following Sec. \ref{sec:tps}, we first write new conjugate variables
in terms of $\c,\r$,\begin{subnumcases}
{\left(\frac{2\pi\M}{\N}\c,\frac{2\pi\L}{\N}\r\right)\rightarrow}
\left(\sqrt{\frac{2\pi}{\N}}\L\x,\sqrt{\frac{2\pi}{\N}}\L\p\right) & $\dfs\rightarrow\ft$ \label{eq:tortopln}\\   
\left(\M\k,\frac{2\pi\L}{\N}\NN\right) &  $\dfs\rightarrow\fs1$ \label{eq:tortocyl1}\\   
\left(\frac{2\pi\M}{\N}\NN,\L\k\right) & $\dfs\rightarrow\fs2 $  \label{eq:tortocyl2}   
\end{subnumcases}where we have set $\M=\L$ for the first case because we do not need
different length scales there. We then take the $\N\rightarrow\infty$
limit and write new Hamiltonians $H_{\ft},H_{\fs1},H_{\fs2}$ which
serve as extensions of $H_{\dfs}$ into $\ft$ and $\fs$; the remainders
of the respective limit-taking procedures are discussed in the next
two subsections.

Previous efforts have rigorously studied similar embeddings in the
past, in particular Barker \cite{Barker2001,Barker2001a,Barker2001b,Barker2003}
and Digernes et al. \cite{DIGERNES1994}. However, the former requires
exact knowledge of the eigenstructure of both $H_{\dfs}$ and $H_{\ft/\fs}$
(see \cite{Barker2001b}, Prop. 3.7) and is thus rigorously applicable
to only simple examples. The latter constrains the Hamiltonian to
be of a different form than $H_{\dfs}$. Here, we extend the procedure
in the Supplement of Ref. \cite{Massar2008} to any $H_{\dfs}$ (in
the language of \cite{Barker2001}, by ``dead reckoning'') with
the goal of creating a meaningful extension of the $\dfs$ system
into $\ft$ and $\fs$. We apply these procedures to six models (see
Table \ref{t:2}), obtaining continuum and rotor generalizations that
in some cases have not been known before. We sometime keep track of
the symmetries of the model to demonstrate that our limits are symmetry-preserving.

\subsection{Continuum limit $\protect\dfs\rightarrow\protect\ft$\label{subsec:Continuum-limit}}

For this case, we take $\N\rightarrow\infty$ and expand around the
center $(0,0)$ of $(\c,\r)$-phase space. We could in principle expand
around a generic point $(s_{0},m_{0})$, but we can always redefine
that to be the origin. This procedure can also be generalized to $K$
$\dfs$ systems $(\c,\r)^{\otimes K}$; we stick to one for simplicity.
First, let us take
\begin{equation}
\frac{\L^{2}}{\N}\rightarrow2\pi\,.\label{eq:limit}
\end{equation}
We perform this limit in order to remove any factors of $\nicefrac{1}{\sqrt{\N}}$
occurring in the expansion of some $f_{k},g_{k}$ later on, noting
that it is not necessary if such factors occur for all $k$. The case
when this step is necessary is only for the Rabi model in Sec. \ref{sec:rabi2}.
We pick $2\pi$ in order to have the $\fs\rightarrow\ft$ limits conveniently
produce the same result as we are about to produce, but this is done
for convention since a sequence of rational numbers can yield any
real. As a sanity check, we see that that $\L=O(\sqrt{\N})$, i.e.,
the hopping between sites (determined by $\L$) does not increase
faster than the total number of sites (proportional to $\N$).

Recalling that we have redefined variables as in Eq. (\ref{eq:tortopln}),
let us approximate $f_{k}(2\pi\x)$ and $g_{k}(2\pi\p)$ with their
expansions around the zero eigenvalue of $\x,\p$, respectively. Such
an expansion can be done in a similar way as operator exponentiation,
i.e., by working in a basis for which the two functions are diagonal
and then expanding each of their eigenvalues. Such an expansion \textit{will
not} hold for arbitrarily high eigenvalues of $H_{\dfs}$ since they
are not always much less than one (e.g., near the maximal values of
$\c,\r$). This means that, as $\N\rightarrow\infty$, we have to
keep projecting ourselves to the intersection of the subspaces of
small eigenvalues of $\x$ and $\p$. Consequently, the eigenstates
of $H_{\dfs}$ which remain in such a limit are only those which are
centered around $(x,p)=(0,0)$ and have small variance in either variable.
Expanding, we obtain (apart from a constant shift in energy)
\begin{equation}
H_{\dfs}\sim H_{\ft}\coloneqq-\sum_{k}A_{k}\p+B_{k}\x+{\textstyle \half}C_{k}\p^{2}+{\textstyle \half}D_{k}\x^{2}+E_{k}\x\p+E_{k}^{\star}\p\x\,,\label{eq:hpln}
\end{equation}
with the coefficients $A_{k},B_{k},C_{k},D_{k},E_{k}$ obvious functions
of $f_{k},g_{k}$ and their derivatives (evaluated at zero). Thus,
such a limit always yields a Hamiltonian consisting of linear and
bilinear terms.

Let us discuss when the $\dfs\rightarrow\ft$ limit corresponds to
a physically meaningful low-energy expansion of $H_{\dfs}$. A trivial
sufficient condition is that the ground state subspace of $H_{\dfs}$
is localized around $(s,m)=(0,0)$. That way, expansion around $(s,m)=(0,0)$
encapsulates the ground state subspace and the low-energy excited
states. If $f_{k}(0)g_{k}(0)$ is a global maximum for each $k$,
then the minus sign in front of the sum (\ref{eq:torgen}) guarantees
that the lowest-energy states will be centered around $(s,m)=(0,0)$.
{[}If there is another maximum at say $f_{k}(\nicefrac{\N}{2})g_{k}(\nicefrac{\N}{2})$
for all $k$, then expansion will of course ignore the ground state
centered at $(\nicefrac{\N}{2},\nicefrac{\N}{2})$.{]} However, being
centered at the origin still does not guarantee that the ground-state
subspace is localized to the same degree in $s$ as it is in $m$.
Examples of systems whose ground states are centered but not equally
localized around $(0,0)$ are $H_{\dfs}=-\cos(\frac{2\pi}{\N}\c)-\cos^{\left\lfloor J\right\rfloor }(\frac{2\pi}{\N}\r)$
and $H_{\dfs}=-\cos(\frac{2\pi}{\N}\c)-J\cos(\frac{2\pi}{\N}\r)$
for $J\gg1$. In both cases, the second term gives a higher energy
penalty for states near the origin than the first term, so expanding
only the second term is more appropriate. A similar example of such
an expansion (albeit of $\fs\rightarrow\ft$ type) is the expansion
of the cosine term in the Josephson junction Hamiltonian, 
\begin{equation}
\alpha\NN^{2}+\beta\cos\k\rightarrow\a\p^{2}+\b\x^{2}\label{eq:jj}
\end{equation}
(where $\alpha,\beta$ are real and $\NN,\k$ are $\fs$ conjugate
variables), to obtain the $\ft$ harmonic oscillator \cite{girvinbook}.
This limit is only valid in the $\beta\gg\alpha$ region of parameter
space since only then the zero-point fluctuations in the phase are
small \textemdash{} the ground state and its lowest-energy excitations
are well-supported by the subspace of eigenstates $\{|\theta\ket\}_{\theta\leq\theta_{\text{max}}}$
of $\k$ with $\theta_{\text{max}}\ll1$. We will not perform such
limits here since they require certain terms to be more dominant than
others; we assume that the contribution of all terms is approximately
equal.

\subsection{Rotor limits $\protect\dfs\rightarrow\protect\fs1,2$\label{subsec:Rotor-limits}}

These limits are considerably simpler than the continuum limit in
that they only involve the hopping length scales $\L,\M$ and there
is no expansion. For $\dfs\rightarrow\fs1$ from Eq. (\ref{eq:tortocyl1}),
we take $\L,\N\rightarrow\infty$ such that 
\begin{equation}
\frac{\L}{\N}\rightarrow\Phi<1\,,
\end{equation}
where $\an$ is irrational. The irrationality of $\Phi$ breaks the
periodicity of $g_{k}$, yielding
\begin{equation}
H_{\dfs}\sim H_{\fs1}\coloneqq-\half\sum_{k}f_{k}\left(\M\k\right)g_{k}\left(2\pi\Phi\NN\right)+H.c.\,.
\end{equation}
For the other limit $\dfs\rightarrow\fs2$ in Eq. (\ref{eq:tortocyl2}),
we let $\nicefrac{\M}{\N}\rightarrow\Phi$, yielding
\begin{equation}
H_{\dfs}\sim H_{\fs2}\coloneqq-\half\sum_{k}f_{k}\left(2\pi\Phi\NN\right)g_{k}\left(\L\k\right)+H.c.\,.
\end{equation}
These two limits are different whenever $f_{k}\neq g_{k}$ for some
$k$. 

There should not be any issues of physical validity with this limit
(other than the fact that the system dimension is now infinite) since
we are not expanding anything. Alternatively, we could expand $g_{k}$
for all $k$, leaving $f_{k}$ intact, given the assumption that low-energy
states are more localized in $\k$ than in $\NN$ (given $\L=1$).
As mentioned in the previous subsection, such a limit also yields
interesting physics, but we do not delve too much on it here.

To complete the diagram (\ref{eq:diagram}), we can take the $\fs1\rightarrow\ft$
limit by letting $\L\rightarrow\infty$ and expanding $f_{k},g_{k}$.
Following Eq. (\ref{eq:thirdlim}), the new variables this time are
$\x\leftarrow\frac{\L}{2\pi}\k$ and $\p\leftarrow\frac{2\pi}{\L}\NN$.
In addition, we take $\Phi\propto\nicefrac{1}{\L}$, where the proportionality
constant can be any real number. We pick $\Phi=\nicefrac{2\pi}{\L}$
so that this expansion reproduces $H_{\ft}$ (\ref{eq:hpln}) within
second order in $\x,\p$. The $\fs2\rightarrow\ft$ limit is performed
the same manner, also yielding $H_{\ft}$.

\section{Harmonic oscillator\label{sec:sho}}

Let us begin with the simple and known \cite{azbel,Hofstadter1976,Barker2000,Barker2000a,Massar2008}
example of $\dfs$ and $\fs$ analogues of the harmonic oscillator
\begin{equation}
H_{\ft}^{\text{sho}}\coloneqq\half\left(\p^{2}+\x^{2}\right)\,.
\end{equation}
First, consider a $\dfs$-type harmonic oscillator\begin{subequations}
\begin{eqnarray}
H_{\dfs}^{\text{sho}} & \coloneqq & -\half\left(\ZZ^{\M}+H.c.\right)-\half\left(\X^{\L}+H.c.\right)\label{eq:hdft0}\\
 & = & -\cos\left(\frac{2\pi\M}{\N}\c\right)-\cos\left(\frac{2\pi\L}{\N}\r\right)\label{eq:hdft}\\
 & = & -\sum_{\ci\in\Z_{\N}}\cos\left(2\pi\frac{\M}{\N}\ci\right)|\ci\ket\bra\ci|-\half\sum_{\ci\in\Z_{\N}}\left(\lket{\ci+\L}\bra\ci|+|\ci\ket\bra\ci+\L|\right)\,,\label{eq:hdft2}
\end{eqnarray}
\end{subequations}where we use $\X$ (\ref{eq:X}) and $\ZZ$ (\ref{eq:Z})
in the first line and the basis $\{|s\ket\}_{s\in\Z_{\N}}$ of eigenstates
of $\c$ (\ref{eq:dfsconj}) in the last. This Hamiltonian corresponds
to a quantum system on a ring with modulated periodic potential and
an $\L$-site hopping term. We can certainly block diagonalize it
into $\L$ blocks if $\N$ is a multiple of $\L$, but we do not concern
ourselves with such special cases. This model is viewed as an analogue
of the oscillator because one can recover $H_{\ft}^{\text{sho}}$
in the $\dfs\rightarrow\ft$ limit. We can already see the $\ft$
structure for $\M=1$ if we recall that $\ZZ+\ZZ^{\dg}$ is, up to
a constant, the discrete Laplacian \cite{DIGERNES1994}; it will become
an ordinary Laplacian in the limit below.

\begin{figure}
\begin{centering}
\includegraphics[width=0.5\columnwidth]{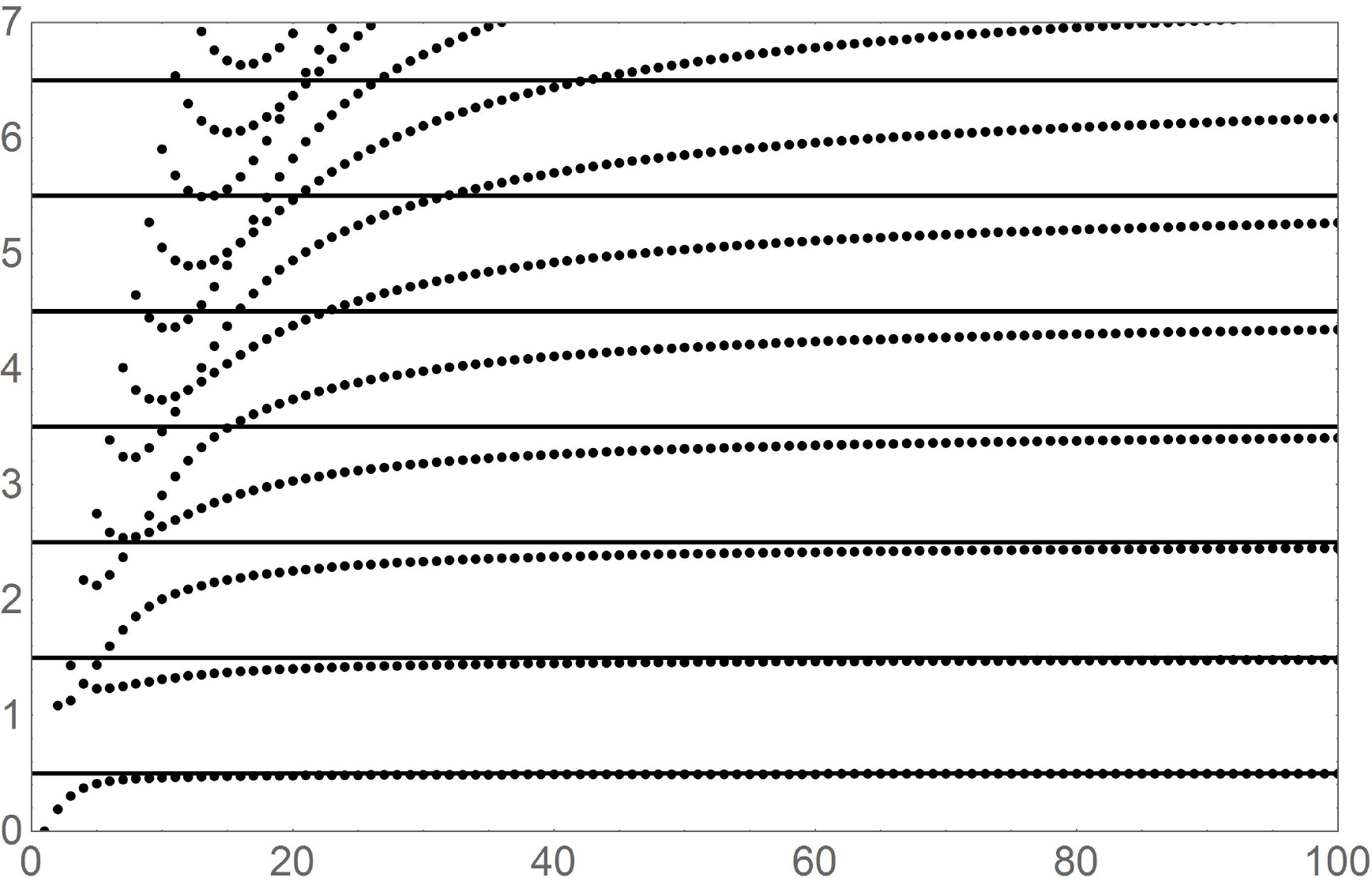}
\par\end{centering}
\caption{\label{f:dft}Eigenvalues of $\frac{\protect\N}{2\pi}(H_{\protect\dfs}^{\text{sho}}-2)$
(\ref{eq:hdft}) vs. $\protect\N$ for $\protect\L=\protect\M=1$.
One can see that they approach true harmonic oscillator eigenvalues
of an integer plus a half (horizontal lines) as $\protect\N\rightarrow\infty$.
This behavior persists for higher values of $\protect\L$.}
\end{figure}

Applying the general technique of Sec. \ref{subsec:Continuum-limit}
\textemdash{} setting $\M=\L$, redefining the conjugate variables
(\ref{eq:tortopln}), expanding around $(s,m)=(0,0)$ in the $\N\rightarrow\infty$
limit, and taking $\nicefrac{\L^{2}}{\N}\rightarrow2\pi$ \textemdash{}
yields
\begin{eqnarray}
H_{\dfs}^{\text{sho}} & \sim & \left(\L\sqrt{\frac{2\pi}{\N}}\right)^{2}\cdot\half\left(\p^{2}+\x^{2}\right)=\left(2\pi\right)^{2}H_{\ft}^{\text{sho}}\,.\label{eq:dfs2ft}
\end{eqnarray}
To check this limit, we can plot the eigenvalues of $\frac{\N}{2\pi}(H_{\dfs}^{\text{sho}}-2)$
for a given $\L$ and with increasing $\N$. As seen in Fig. \ref{f:dft},
the eigenvalues approach those of the continuum harmonic oscillator.
In fact, this limit has been proven to yield $H_{\ft}^{\text{sho}}$
exactly (see example C.4 in Ref. \cite{Barker2001b} and references
therein), so we are certain that the $\dfs\rightarrow\ft$ limit is
physically meaningful in this case. It should not be surprising since
the ground state of $H_{\dfs}^{\text{sho}}$ is localized around $(s,m)=(0,0)$
and expansion of both cosines adds the lowest possible energy penalty.
Such a limit is also generalizable to $\M\neq\L$, allowing $H_{\ft}^{\text{sho}}$
to have a free frequency parameter.

Now let us take $H_{\dfs}^{\text{sho}}$ to one of two $\fs$-type
harmonic oscillators. Following Sec. \ref{subsec:Rotor-limits}, redefine
variables as in Eq. (\ref{eq:tortocyl1}) and let $\nicefrac{\L}{\N}\rightarrow\an$,
yielding\begin{subequations}
\begin{eqnarray}
H_{\fs1}^{\text{sho}} & \coloneqq & -\cos(\M\k)-\cos\left(2\pi\an\NN\right)\label{eq:hfs2}\\
 & = & -\half\sum_{N\in\Z}\left(\lket{N+\M}\bra N|+|N\ket\bra N+\M|\right)-\sum_{N\in\Z}\cos\left(2\pi\an N\right)|N\ket\bra N|\,,
\end{eqnarray}
\end{subequations}where $\an<1$ is a positive irrational. In the
context of the quantum Hall effect, $H_{\fs1}^{\text{sho}}$ models
an electron on a 2D lattice in the presence of a magnetic field and
$\Phi$ is the magnetic flux per unit cell. In that context, $H_{\dfs}^{\text{sho}}$
is called Harper's equation \cite{wang1987} and $H_{\fs1}^{\text{sho}}$
the almost Mathieu operator. In particular, $H_{\fs1}^{\text{sho}}$
with $\Phi=\half(\sqrt{5}-1)$ corresponds to the Fibonacci quasicrystal
\cite{Kraus2012a}. There is of course another way to obtain an $\fs$-type
Hamiltonian from $H_{\dfs}^{\text{sho}}$ by following Eq. (\ref{eq:tortocyl2})
and letting $\nicefrac{\M}{\N}\rightarrow\an$, yielding the $\fs2$
Hamiltonian\begin{subequations}
\begin{eqnarray}
H_{\fs2}^{\text{sho}} & \coloneqq & -\cos\left(2\pi\an\NN\right)-\cos(\L\k)\,.\label{eq:hfs1}
\end{eqnarray}
\end{subequations}In this case, the two $\fs$ limits yield the same
result.

Finishing off with the $\fs\rightarrow\ft$ limit, let $\an\rightarrow\nicefrac{2\pi}{\L}\rightarrow0$
and expand both cosine terms in $H_{\fs1}^{\text{sho}},H_{\fs2}^{\text{sho}}$.
In the aforementioned context of the quantum Hall effect, this limit
is related to the quasiclassical limit of vanishing field and recovers
the famous Landau-level problem \cite{wang1987} \textemdash{} a simple
harmonic oscillator,
\begin{equation}
H_{\fs1}^{\text{sho}}\sim\frac{\left(\Phi\L\right)^{2}}{2}\p^{2}+\frac{\left(2\pi\right)^{2}}{2}\x^{2}=\left(2\pi\right)^{2}H_{\ft}^{\text{sho}}\,.
\end{equation}
The result is thus the same as the direct $\dfs\rightarrow\ft$ procedure
from Eq. (\ref{eq:dfs2ft}).

\section{Baxter $Z_{\protect\N}$ parafermionic spin chain\label{sec:Baxter}}

Using Eqs. (\ref{eq:X}-\ref{eq:Z}) for the Weyl operators $\X$
and $\ZZ$, the Baxter $\Z_{\N}$ spin chain Hamiltonian \cite{Baxter1989,Fendley2014,Alcaraz2017}
(in Hermitian form) reads\begin{subequations}
\begin{eqnarray}
H_{\dfs}^{\text{bax}} & \coloneqq & -\frac{\O}{2}\sum_{k=1}^{K}\left(\ZZ_{k}^{\M}+H.c.\right)-\frac{g}{2}\sum_{k=1}^{K-1}\left(\X_{k}^{\L}\X_{k+1}^{-\L}+H.c.\right)\label{eq:isingtor}\\
 & = & -\O\sum_{k=1}^{K}\cos\left(\frac{2\pi\M}{\N}\c_{k}\right)-g\sum_{k=1}^{K-1}\cos\left(\frac{2\pi\L}{\N}\left[\r_{k+1}-\r_{k}\right]\right)\,,\label{eq:baxtercosine}
\end{eqnarray}
\end{subequations}where the parameters $\O,g$ are of the same order
of magnitude (so that we can expand both cosines). The $\N=2$ case
reduces to the original Ising model.\footnote{Another generalization of the Ising model \textemdash{} the quantum
Potts (e.g., \cite{Solyom1981,Cant1983,Fendley2012}) or quantum clock
\cite{Vaezi2013} model \textemdash{} is not amenable to our limit-taking
procedures because it contains a sum over all $\L,\M$ and so restricts
us from having $\L,\M=O(\sqrt{\N})$ for large $\N$. } For general $\N$, each site $k$ corresponds lives in its own $\dfs$
phase space, and the full model is therefore in $\dfs^{\otimes K}$.
One can also add phases (e.g., $\ZZ_{k}\rightarrow\ZZ_{k}e^{i\phi}$)
\cite{Fendley2012}, leading to more complicated behavior (similar
to the chiral version of the Rabi model in Sec. \ref{sec:rabi2}).
This model is equivalent to a parafermion chain via a nonlocal extension
of the Jordan-Wigner transformation \cite{Moran2017}. When written
in terms of the conjugate variables $\{\c,\r\}^{\otimes K}$, the
model's (\ref{eq:baxtercosine}) invariance upon the collective translations
$\r_{k}\rightarrow\r_{k}+1$ and reflections $\r_{k}\rightarrow-\r_{k}$
(both for all $k$) are made bare. The operator for the former symmetry
is simply $\ZZ^{\otimes K}$; we will keep track of this symmetry
as the systems travels to the $\ft$ and $\fs$ phase spaces.

We now apply the limit-taking procedures from Sec. \ref{sec:Continuum-and-rotor}
to $H_{\dfs}^{\text{sho}}$, obtaining its $\fs$ and $\ft$ limits.
If we take $H_{\dfs}^{\text{sho}}$ (\ref{eq:hdft}) to be the canonical
harmonic oscillator in $\dfs$ phase space, then we can see that $H_{\dfs}^{\text{bax}}$
(\ref{eq:baxtercosine}) is nothing but a chain of coupled $\dfs$
oscillators. It should thus not come as a surprise that the $\dfs\rightarrow\ft$
limit, as we shall see, produces a chain of $\ft$ oscillators. Up
to a constant offset and a multiplicative factor of $(2\pi)^{2}$,
redefining variables per Eq. (\ref{eq:tortopln}) and expanding the
cosines around $(s,m)=(0,0)$ yields
\begin{equation}
H_{\ft}^{\text{bax}}\coloneqq\frac{\O}{2}\sum_{k=1}^{K}\p_{k}^{2}+\frac{g}{2}\sum_{k=1}^{K-1}\left(\x_{k+1}-\x_{k}\right)^{2}\,.\label{eq:isingpln}
\end{equation}
The aforementioned symmetries clearly survive: $H_{\ft}^{\text{bax}}$
is invariant under global reflection $\x_{k}\rightarrow-\x_{k}$ and
any global shifts $\x_{k}\rightarrow\x_{k}+\xi$ for all $k$ and
any real $\xi$. It will interesting to see whether the spectrum of
this simple model can be rigorously obtained in the large-$\N$ limit
of the exact spectrum of the Baxter model, as was done for the oscillator
\cite{Barker2001b}.

Let us now turn the $\dfs$ phase spaces of $H_{\dfs}^{\text{bax}}$
into infinite ladders ($\fs1$), taking the $\c,\r\rightarrow\NN,\k$
limit from Sec. \ref{subsec:Rotor-limits}, yielding
\begin{eqnarray}
H_{\fs1}^{\text{bax}} & \coloneqq & -\O\sum_{k=1}^{K}\cos\left(2\pi\an\NN_{k}\right)-g\sum_{k=1}^{K-1}\cos\left(\L\left[\k_{k+1}-\k_{k}\right]\right)\,.
\end{eqnarray}
Expanding the first cosine in a limit ($\O\gg g$) similar to that
in eq. (\ref{eq:jj}) immediately yields a Hamiltonian for an array
of coupled Josephson Junctions in the quantum regime \textemdash{}
the quantum XY model (e.g., \cite{Peskin1978,Capriotti2004}). Alternatively,
we can take the infinitely dense circle limit ($\fs2$), in which
$\c,\r\rightarrow\k,\NN$:
\begin{equation}
H_{\fs2}^{\text{bax}}\coloneqq-\O\sum_{k=1}^{K}\cos\left(\M\k_{k}\right)-g\sum_{k=1}^{K-1}\cos\left(2\pi\an\left[\NN_{k+1}-\NN_{k}\right]\right)\,.
\end{equation}
This corresponds to an infinite ladder of sites once more, but this
time there is a hopping term and an interaction that is diagonal in
the $\NN$ basis. Expanding the latter in a procedure similar to that
in eq. (\ref{eq:jj}) yields a Hubbard model of sorts, with interactions
of the form $(\NN_{k+1}-\NN_{k})^{2}$. In both $\fs$ versions, invariance
under global reflection and shifts of all $\k_{k}$ ($\NN_{k}$) in
the case $\fs1$ ($\fs2$) is preserved.

Both the $\fs1$ and $\fs2$ cases can then be taken to the $\ft$
case by performing the procedures described in Sec. \ref{sec:sho}
for the simple harmonic oscillator. For $H_{\fs1}^{\text{bax}}$,
take $\Phi=\nicefrac{2\pi}{\L}$, $\frac{\L}{2\pi}\k\rightarrow\x$,
$\frac{2\pi}{\L}\NN\rightarrow\p$, and $\L\rightarrow\infty$. For
$H_{\fs2}^{\text{bax}}$, take $\Phi=\nicefrac{2\pi}{\M}$, $\frac{\M}{2\pi}\k\rightarrow\p$,
$\frac{2\pi}{\M}\NN\rightarrow\x$, and $\M\rightarrow\infty$. Both
cases reduce those Hamiltonians to $H_{\ft}^{\text{bax}}$ (\ref{eq:isingpln}).
These two limits are the reason we introduced both length scales $\M$
and $\L$ in $H_{\dfs}^{\text{bax}}$ (\ref{eq:isingtor}); performing
the $\dfs\rightarrow\ft$ procedure alone does not require them.

\section{$\protect\N$-state Rabi model\label{sec:rabi2}}

The original quantum Rabi model Hamiltonian \cite{rabi,rabi2} consists
of an interacting two-level system (i.e., a qubit) and a harmonic
oscillator and is arguably one of the simplest non-linear, non-trivial
models. Letting $\s_{x},\s_{z}$ be the Pauli matrices of the qubit
and $\bb/\bb^{\dg}$ be the lowering/raising operators of the oscillator,
the Rabi Hamiltonian is
\begin{equation}
\frac{1}{\hbar}H_{\N=2}^{\text{rabi}}\coloneqq\o\left(\bb^{\dg}\bb+\half\right)-\O\s_{z}+g\left(\bb+\bb^{\dg}\right)\s_{x}\,.
\end{equation}
The three real parameters are the positive oscillator frequency $\o$,
the qubit Larmor frequency $\O$, and the qubit-oscillator coupling
$g$. This model and its close relatives have been used in numerous
contexts to simulate qubit-oscillator systems, including:
\begin{enumerate}
\item A two-level atom coupled to a cavity electromagnetic field in quantum
optics and superconducting circuits. In this context, the model is
a precursor to the Jaynes-Cummings model \cite{jc, Paul1963, Shore1993}:
ignoring the term $\bb\s_{-}$ and its conjugate (the rotating-wave
approximation) yields that famous Hamiltonian.
\item A spinful electron coupled to a magnetic field; the related model
is called the Landau level problem with Dresselhaus or Rashba spin-orbit
coupling \cite{lewenstein2006,tomka2015,bernevig2006}.
\item Ultra-cold alkali atoms in a magnetic field and double-well potential
\cite{Deutsch2000}.
\item An exciton interacting with lattice vibrations at two sites of a crystalline
system, where the model is called the single-mode spin boson or the
two-site Holstein model \cite{spin_boson_shore_sander_fg,wagner}.
\item Strong magnetic coupling between an NV center and a nanomechanical
oscillator in a longitudinal \cite{rabl2009} or transverse \cite{macquarrie2013}
field.
\end{enumerate}
Unlike the Jaynes-Cummings model, this model has a $\Z_{2}$ symmetry.
Namely, the Hamiltonian commutes with 
\begin{equation}
\V_{\N=2}=\left(-1\right)^{\bb^{\dg}\bb}\s_{z}\,,
\end{equation}
which squares to identity and represents a joint parity of the qubit
and the oscillator {[}recall that $\F_{\ft}^{2}=\left(-1\right)^{\bb^{\dg}\bb}$
(\ref{eq:parity-1}){]}. The model is also real, so $H_{\N=2}^{\text{rabi}}$
commutes with the complex conjugation operator $K$.

The $\N$-state Rabi model \cite{pub009} is an extension of the qubit
($\N=2$) Rabi model to qudits that naturally extends the symmetry
of the qubit case:\begin{subequations}
\begin{eqnarray}
\frac{1}{\hbar}H_{\dfs}^{\text{rabi}} & \coloneqq & \o\left(\bb^{\dg}\bb+\half\right)-\frac{\O}{2}\left(\ZZ^{\M}+H.c.\right)+g\left(\bb\X^{\L}+H.c.\right)\label{eq:rabi}\\
 & = & \o\left(\bb^{\dg}\bb+\half\right)-\O\cos\left(\frac{2\pi\M}{\N}\c\right)+g\left(\bb e^{-i\frac{2\pi\L}{\N}\r}+H.c.\right)\label{eq:rabi2}
\end{eqnarray}
\end{subequations}(Note how the $\N=2$ case reduces to the original
Rabi model.) Originally, the $\N$-state Rabi model included more
couplings using other powers of $\ZZ$, but we omit those for simplicity.
The $\Z_{2}$ symmetry from the $\N=2$ case generalizes to a $\Z_{\N}$
symmetry with generator
\begin{equation}
\V=e^{i\frac{2\pi}{\N}\L\,\bb^{\dg}\bb}\ZZ=\exp\left[i\frac{2\pi}{\N}\left(\L\,\bb^{\dg}\bb+\c\right)\right]\,.
\end{equation}
We further say here that the model actually has a \textit{dihedral}
symmetry. In addition to $\V$, the Hamiltonian is invariant under
the operation $\c\rightarrow-\c$. This reflection around the $\r=0$
axis is represented by the antiunitary operator 
\begin{equation}
\U\coloneqq\R K\,,
\end{equation}
where $K$ is complex conjugation and $\R$ (\ref{eq:parity}) is
the $\dfs$ parity operator taking $\c,\r\rightarrow-\c,-\r$ modulo
$\N$. The full symmetry is then the dihedral group $D_{2\N}=\Z_{\N}\rtimes\mathbb{Z}_{2}$,
where the $\Z_{\N}$ piece is generated by $\V$ while the $\Z_{2}$
piece is generated by $\U$. It is interesting to note that one can
reduce the $\Z_{\N}\rtimes\mathbb{Z}_{2}$ symmetry to $\Z_{\N}$
by giving the cosine term a phase. In such a \textit{chiral} version
of the $\dfs$-type Rabi model \cite{Zhang2014}, $\c\rightarrow-\c$
will no longer be a symmetry. Observing Eq. (\ref{eq:rabi2}), we
can see that the qudit consists of a ring of $\N$ sites with energies
$-\O\cos\left(2\pi\frac{\M}{\N}\c\right)$ and that the hopping term
$\bb e^{-i\frac{2\pi}{\N}\L\r}$ takes the qudit state $|\ci\ket$
to the state $|\ci+\L\ket$ while also absorbing a photon. The conjugate
term corresponds to emitting a photon and moving $\L$ sites in the
other direction along the qudit ring. 

Let us perform the $\dfs\rightarrow\ft$ limit from Sec. \ref{subsec:Continuum-limit},
which yields a $\ft$ phase space for the degrees of freedom of the
$\N$-state system. Just like in for previous two models, we redefine
variables according to Eq. (\ref{eq:tortopln}) and expand around
the origin $(s,m)=(0,0)$ of $\dfs$ phase space. To our knowledge,
the $\dfs\rightarrow\ft$ limit is different from previous limits
of the Rabi model, e.g., the large spin limit \cite{Ghose2005,Bakemeier2012,gilmorebook}.
Letting $\yy\coloneqq\frac{1}{\sqrt{2}}(\bb+\bb^{\dg})$ and $\P=-i\frac{1}{\sqrt{2}}(\bb-\bb^{\dg})$
(with $[\yy,\P]=i$) be the original oscillator degrees of freedom
(which remain unchanged), this yields
\begin{eqnarray}
\frac{1}{\hbar}H_{\ft}^{\text{rabi}} & \coloneqq & \frac{\o}{2}\left(\P^{2}+\yy^{2}\right)+2\O\pi^{2}\p^{2}+\sqrt{2}g\left(2\pi\x\P+\yy\left[\id-2\pi^{2}\x^{2}\right]\right)-\O\,.\label{eq:rabift}
\end{eqnarray}
Displacing that oscillator by $-\frac{\sqrt{2}g}{\o}$ {[}via the
operator $\D$, which takes $\yy\rightarrow\yy-\frac{g}{\o}$ and
same with $\P${]} gets rid of the linear $\yy$ term and yields
\begin{eqnarray}
\frac{1}{\hbar}\D H_{\ft}^{\text{rabi}}\D^{\dg} & = & \frac{\o}{2}\left(\P^{2}+\yy^{2}\right)+2\pi^{2}\left(\O\p^{2}+\frac{2g^{2}}{\o}\x^{2}\right)+2\sqrt{2}g\pi\x\left(\P-\pi\x\yy\right)-\left(\O+\frac{2g^{2}}{\o}\right)\,.
\end{eqnarray}
We thus see how the Rabi model reduces to a coupled oscillator model
in the $\dfs\rightarrow\ft$ limit. We can see that what is obtained
is none other than the optomechanical Hamiltonian of a mechanical
mode (which used to the $\N$-state system) interacting with a light
mode via the $\x^{2}\bb$ coupling. The $\s_{+}\bb$ coupling induced
by light on a two-level atom has become, in this limit, radiation
pressure on an oscillator.

Performing the $\dfs\rightarrow\fs1,2$ limits from Sec. \ref{subsec:Rotor-limits}
yields the respective coupled-rotor systems
\begin{eqnarray}
\frac{1}{\hbar}H_{\fs1}^{\text{rabi}} & \coloneqq & \o\left(\bb^{\dg}\bb+\half\right)-\O\cos\left(2\pi\an\NN\right)+g\left(\bb e^{-i\L\k}+H.c.\right)\\
\frac{1}{\hbar}H_{\fs2}^{\text{rabi}} & \coloneqq & \o\left(\bb^{\dg}\bb+\half\right)-\O\cos\left(\M\k\right)+g\left(\bb e^{-i2\pi\an\NN}+H.c.\right)\,.
\end{eqnarray}
The first model corresponds to an infinite ladder consisting of sites
$|N\ket$ in a $\O$-mediated quasiperiodic potential and with the
coupling term causing a particle on a site to move either to the left
or right, depending on whether the particle absorbs or emits a photon.
The second model corresponds to an infinite ladder of sites with hopping
strength $\O$, but this time each site $|N\ket$ is linearly coupled
to an oscillator with strength $g$ and quasiperiodic phase $2\pi\Phi N$.
Naturally, the dihedral ($D_{2\N}=\Z_{\N}\rtimes\mathbb{Z}_{2}$)
symmetry of the $\dfs$-type Rabi model is extended to a $U(1)\rtimes\mathbb{Z}_{2}=O(2)$
symmetry. The $U(1)$ piece is generated by $\L\bb^{\dg}\bb+\NN$
for $\fs1$ and by $2\pi\Phi\bb^{\dg}\bb+\k$ for $\fs2$. The $\Z_{2}$
piece just corresponds to reflection: $\NN\rightarrow-\NN$ for $\fs1$
and $\k\rightarrow-\k$ for $\fs2$. Both $H_{\fs1}^{\text{rabi}}$
and $H_{\fs2}^{\text{rabi}}$ can then be taken to $H_{\ft}^{\text{rabi}}$
(\ref{eq:rabift}) via the procedures described in Sec. \ref{subsec:Rotor-limits}.

\section{Kitaev toric code\label{sec:toric-code}}

The toric code \cite{Kitaev2003} was proposed as a simple model for
topological quantum computation. The degrees of freedom are qubits
typically living on vertices of a square lattice with periodic boundary
conditions (i.e., a torus). The Hamiltonian is written as a sum of
products of four Weyl operators. We denote operators $O$ acting nontrivially
on a site ``$\bullet$'' on a plaquette ``$\square$'' or vertex/star
``$+$'' as, e.g., $O_{\upl}$ or $O_{\uvr}$, respectively. The
$\Z_{\N}$ toric code Hamiltonian \cite{Kitaev2003,Wen2003,Bullock2007}
is then\begin{subequations}
\begin{eqnarray}
H_{\dfs}^{\text{tor}} & \coloneqq & -\frac{J_{z}}{2}\sum_{\square}\left(\ZZ_{\dpl}^{\M}\ZZ_{\rpl}^{\M}\ZZ_{\upl}^{-\M}\ZZ_{\lpl}^{-\M}+H.c.\right)-\frac{J_{x}}{2}\sum_{+}\left(\X_{\rvr}^{\L}\X_{\uvr}^{\L}\X_{\lvr}^{-\L}\X_{\dvr}^{-\L}+H.c.\right)\\
 & = & -J_{z}\sum_{\square}\cos\left(\frac{2\pi\M}{\N}\c_{\square}\right)-J_{x}\sum_{+}\cos\left(\frac{2\pi\L}{\N}\r_{+}\right)\,,
\end{eqnarray}
\end{subequations}where $J_{x,z}>0$ and $0<\M,\L<\N$ are the hopping
length scales for the respective variables. The joint degrees of freedom
associated with \textit{going around} a plaquette $\square$ counterclockwise
and \textit{going out }of a vertex $+$, respectively, are
\begin{equation}
\c_{\square}=\c_{\dpl}+\c_{\rpl}-\c_{\upl}-\c_{\lpl}\,\,\,\,\,\,\,\,\,\,\,\,\,\,\,\,\,\,\text{and}\,\,\,\,\,\,\,\,\,\,\,\,\,\,\,\,\,\,\r_{+}=\r_{\rvr}+\r_{\uvr}-\r_{\lvr}-\r_{\dvr}\,.
\end{equation}
(Note how the $\N=2$ case reduces to the original toric code.) The
minus signs in $\c_{\square},\r_{+}$ give the model an orientation
\cite{Viyuela2012a} and make sure that all terms in $H_{\dfs}^{\text{tor}}$
commute with each other, meaning that $H_{\dfs}^{\text{tor}}$ is
\textit{frustration-free}. In the language of $\Z_{\N}$ gauge theory
\cite{Kitaev2003}, $\c_{\square}$ is the magnetic field through
$\square$ while $\r_{+}$ is the electric charge at $+$.

Fixing $\M=\L=1$ for this paragraph, the ground state is one for
which $\c_{\square}=\r_{+}=0$ (modulo $\N$) for all $\square$ and
$+$, respectively, and its lowest-energy excitations have $\c_{\square}=\pm1$
or $\r_{+}=\pm1$ for one $\square$ or $+$. There exist four types
of string-like conserved quantities which determine the ground-state
degeneracy,\begin{subequations}
\begin{eqnarray}
\ZZ_{\leftrightarrow} & = & \cdots\,\ZZ_{\lvr}\ZZ_{\rvr}\cdots,\,\,\,\,\,\,\,\,\,\,\,\,\,\,\,\,\,\,\,\,\,\ZZ_{\updownarrow}=\cdots\,\ZZ_{\dvr}\ZZ_{\uvr}\cdots,\,\,\,\,\,\,\,\,\,\,\,\,\,\,\,\,\,\,\,\,\,\X_{\leftrightarrow}=\cdots\,\X_{\lpl}\X_{\rpl}\cdots,\,\,\,\,\,\,\,\,\,\,\,\,\,\,\,\,\,\,\,\,\,\X_{\updownarrow}=\cdots\,\X_{\dpl}\X_{\upl}\cdots\,,
\end{eqnarray}
\end{subequations}where the product is over all sites along either
one of the two noncontractible loops of the torus. These satisfy $\X_{\leftrightarrow}\ZZ_{\updownarrow}\X_{\leftrightarrow}^{-1}\ZZ_{\updownarrow}^{-1}=\X_{\updownarrow}\ZZ_{\leftrightarrow}\X_{\updownarrow}^{-1}\ZZ_{\leftrightarrow}^{-1}=e^{-i\frac{2\pi}{\N}}$,
implying that there is an $\N^{2}$-dimensional ground-state subspace
that can be characterized by eigenvalues of
\begin{equation}
\c_{\leftrightarrow}=\cdots+\,\c_{\lvr}+\c_{\rvr}+\cdots\mod\N\,\,\,\,\,\,\,\,\,\,\,\,\,\,\,\,\,\,\,\,\text{and}\,\,\,\,\,\,\,\,\,\,\,\,\,\,\,\,\,\,\,\,\c_{\updownarrow}=\cdots+\c_{\dvr}+\c_{\uvr}+\cdots\mod\N\,.\label{eq:degen}
\end{equation}

Once again, we have introduced the $\M,\L$ degrees of freedom only
to perform the $\dfs\rightarrow\fs$ and $\fs\rightarrow\ft$ limits;
they are not required for the direct $\dfs\rightarrow\ft$ limit.
Assuming $J_{x}\approx J_{z}$ and performing the $\dfs\rightarrow\ft$
limit from Sec. \ref{subsec:Continuum-limit} yields, up to constant
offsets and factors, the coupled-oscillator Hamiltonian
\begin{eqnarray}
H_{\ft}^{\text{tor}} & \coloneqq & \frac{J_{z}}{2}\sum_{\square}\p_{\square}^{2}+\frac{J_{x}}{2}\sum_{+}\x_{+}^{2}\,,
\end{eqnarray}
where $\p_{\square}=\p_{\dpl}+\p_{\rpl}-\p_{\upl}-\p_{\lpl}$ and
$\x_{+}=\x_{\rvr}+\x_{\uvr}-\x_{\lvr}-\x_{\dvr}$. This is exactly
the Hamiltonian whose ground states are those of the CV surface code
\cite{Zhang2008}. While this Hamiltonian was known before \cite{Demarie2014},
the $\dfs\rightarrow\ft$ limit-taking procedure provides its direct
derivation from the original toric code. This system has an infinite-dimensional
degeneracy {[}$\c\rightarrow\p$ in Eq. (\ref{eq:degen}){]} and is
gapless.

While the direct $\dfs\rightarrow\ft$ limit reproduces a known instance
of the toric code, taking the $\fs$ detour introduces two new generalizations.
Performing the $\dfs\rightarrow\fs1,2$ limits from Sec. \ref{subsec:Rotor-limits}
yields the respective coupled-rotor systems
\begin{eqnarray}
H_{\fs1}^{\text{tor}} & \coloneqq & -J_{z}\sum_{\square}\cos\left(2\pi\Phi\NN_{\square}\right)-J_{x}\sum_{+}\cos\left(\L\k_{+}\right)\label{eq:lgt}\\
H_{\fs2}^{\text{tor}} & \coloneqq & -J_{z}\sum_{\square}\cos\left(\M\k_{\square}\right)-J_{x}\sum_{+}\cos\left(2\pi\Phi\NN_{+}\right)\,.
\end{eqnarray}
The plaquette and vertex structure is of course preserved, but now
the degree of freedom on each site is a rotor ($\fs$). The degrees
of freedom of these models resemble those of compact $U(1)$ lattice
gauge theory (LGT): $\k_{\square}$ is the magnetic flux term $\Phi$
in the equation below Eq. (6.4.13) in Ref. \cite{wenbook} while $\NN_{+}$
can be interpreted as the electric charge at the center of the ``$+$''.
In fact, the relation between $H_{\fs}^{\text{tor}}$ and $U(1)$
LGT is the same as that between $H_{\dfs}^{\text{tor}}$ and $\Z_{\N}$
LGT \cite{Kitaev2003}: the $H^{\text{tor}}$ systems consist of the
flux term from their corresponding LGT along with a term which represents
a local gauge transformation and whose value is constant within the
ground-state subspace. Given that $U(1)$ gauge theory is always confined
in two dimensions, it remains to be seen whether the rotor models
can admit novel deconfined phases.

\section{Haah cubic code\label{sec:Haah-cubic-code}}

The cubic code \cite{Haah2011} is a three-dimensional generalization
of the toric code and is an example of fracton topological order \cite{Vijay2016}.
It is defined on a cubic lattice, where each site contains \textit{two}
$\dfs(\N=2)$ subsystems. We denote the lattice cubes by~ $\cube$
and operators $O$ acting nontrivially on the first (``$\bullet$'')
or second (``$\times$'') subsystem on a given cube as, e.g., $O_{\ozz}$
or $O_{\tozz}$. The original Hamiltonian is

\begin{align}
H_{\N=2}^{\text{cub}} & \coloneqq-J_{z}\sum_{\cube}A_{\,\,\,\cube}-J_{x}\sum_{\cube}B_{\,\,\,\cube}\,,
\end{align}
where $J_{A,B}>0$ and the sum is over all cubes in the lattice. The
operators $A_{\,\,\,\cube},B_{\,\,\,\cube}$ act nontrivially on 8
out of the 16 sites of the cube; we will not define them for conciseness
and instead directly write the generalized $\dfs(\N\geq2)$ version.
Using the same tricks as for the toric code regarding defining an
orientation and making sure all cubes commute \cite{Viyuela2012a},
one can come up with the model
\begin{align}
H_{\dfs}^{\text{cub}} & \coloneqq-\frac{J_{z}}{2}\sum_{\cube}\left(\ZZ_{\ozz}^{\M}\ZZ_{\zoz}^{\M}\ZZ_{\ooo}^{\M}\ZZ_{\ooz}^{\M}\ZZ_{\tzzz}^{\M}\ZZ_{\tozo}^{\M}\ZZ_{\tzoo}^{\M}\ZZ_{\tooz}^{\M}+H.c.\right)-\frac{J_{x}}{2}\sum_{\cube}\left(\X_{\ozz}^{\L}\X_{\zoz}^{\L}\X_{\ooo}^{\L}\X_{\zzo}^{\L}\X_{\tzzz}^{-\L}\X_{\tozo}^{-\L}\X_{\tzoo}^{-\L}\X_{\tzzo}^{-\L}+H.c.\right)\nonumber \\
 & =-J_{z}\sum_{\cube}\cos\left(\frac{2\pi\M}{\N}\left[\c_{\sdots}+\c_{\scros}\right]\right)-J_{x}\sum_{\cube}\cos\left(\frac{2\pi\L}{\N}\left[\r_{\mdots}-\r_{\mcros}\right]\right)\,,
\end{align}
where the first line is written in terms of Weyl operators (\ref{eq:X}-\ref{eq:Z})
and the second in terms of their corresponding conjugate variables
$\c,\r$ (\ref{eq:dfsconj}). The composite variables are, e.g., $\c_{\sdots}=\c_{\ozz}+\c_{\zoz}+\c_{\ooo}+\c_{\ooz}$,
with the remaining three defined similarly. The relative minus in
the $J_{x}$-term is so that $H_{\dfs}^{\text{cub}}$ is frustration
free. To verify this, one needs make sure that a given cube commutes
with all of the 26 neighboring cubes with which it shares faces, sides,
and corners. We further assume $J_{x}\approx J_{z}$ in order to further
justify the $\dfs\rightarrow\ft$ procedure. A simpler model can be
defined with $\M,\L=1$, but we do not assume this in order to more
conveniently take all of the desired limits.

Performing the $\dfs\rightarrow\ft$ limit from Sec. \ref{subsec:Continuum-limit}
yields, up to constant offsets and factors, the coupled-oscillator
Hamiltonian
\begin{equation}
H_{\ft}^{\text{cub}}\coloneqq\frac{1}{2}\sum_{\cube}\left(\p_{\sdots}+\p_{\scros}\right)^{2}+\frac{1}{2}\sum_{\cube}\left(\x_{\mdots}-\x_{\mcros}\right)^{2}\,.
\end{equation}
Performing the $\dfs\rightarrow\fs1,2$ limits from Sec. \ref{subsec:Rotor-limits}
yields the respective coupled-rotor systems
\begin{align}
H_{\fs1}^{\text{cub}} & \coloneqq-\sum_{\cube}\cos\left(2\pi\Phi\left[\NN_{\sdots}+\NN_{\scros}\right]\right)-\sum_{\cube}\cos\left(\L\left[\k_{\mdots}-\k_{\mcros}\right]\right)\\
H_{\fs2}^{\text{cub}} & \coloneqq-\sum_{\cube}\cos\left(\M\left[\k_{\sdots}+\k_{\scros}\right]\right)-\sum_{\cube}\cos\left(2\pi\Phi\left[\NN_{\mdots}-\NN_{\mcros}\right]\right)\,.
\end{align}
Note that the $\Phi\ll1$ version of this model, in which one of the
cosines is expanded, was independently written down and studied by
Haah \cite{haahtalk}. Since $U(1)$ lattice gauge theory can be deconfined
in three dimensions \cite{wenbook}, it will be interesting to examine
whether these models can also admit interesting deconfined phases.
All Hamiltonians remain frustration-free as that algebraic structure
is preserved in this limit. We have thus generalized the $\N=2$ cubic
code to arbitrary $\N$ as well as to the $\fs$ and $\ft$ phase
spaces. Moreover, this recipe can be applied to any qudit stabilizer
code.

\section{Kitaev honeycomb model\label{sec:Kitaev-model}}

The Kitaev honeycomb model \cite{Kitaev2006} is a two-dimensional
exactly solvable model with an extensive number of conserved quantities
that, along with its relative the toric code (see Sec. \ref{sec:toric-code}),
is a paradigmatic model lying at the intersection of topological quantum
phases of matter and quantum computation. It was originally defined
on a honeycomb lattice; we denote the lattice plaquettes by~ $\hex$
and operators $O$ on a given site ``$\bullet$'' on a plaquette
as, e.g., $O_{\hone}$ or $O_{\hfou}$. The original model is defined
on a set of $\dfs(\N=2)$ sites, but has been extended to lattices
consisting of sites of the type $\dfs$ for arbitrary $\N$ \cite{Barkeshli2015,Fendley2012}
(see \cite{Vaezi2014a} for a different extension). We write such
a model below in terms of Weyl operators (\ref{eq:X}-\ref{eq:Z}),
using the convention of Ref. \cite{Barkeshli2015} but adding the
hopping length scales $\M,\L$ in order to more conveniently take
all of the limits we desire:
\begin{eqnarray}
H_{\dfs}^{\text{hon}} & = & -\frac{1}{2}\sum_{\hex}\left(J_{x}\X_{\hone}^{\L}\X_{\htwo}^{\L}+J_{y}\Y_{\htwo}^{\M,\L}\Y_{\hthr}^{\M,\L}+J_{z}\ZZ_{\hthr}^{\M}\ZZ_{\hfou}^{\M}+H.c.\right)\,,
\end{eqnarray}
where $J_{x,y,z}>0$, $\Y=\ZZ^{-1}\X^{-1}$, and $\Y^{\M,\L}=\ZZ^{-\M}\X^{-\L}$.
In terms of the conjugate variables $\c,\r$ (\ref{eq:dfsconj}),
we have
\begin{eqnarray}
H_{\dfs}^{\text{hon}} & = & -J_{x}\sum_{\hex}\cos\left(\frac{2\pi\L}{\N}\left[\r_{\hone}+\r_{\htwo}\right]\right)-J_{z}\sum_{\hex}\cos\left(\frac{2\pi\M}{\N}\left[\c_{\hthr}+\c_{\hfou}\right]\right)\\
 &  & -J_{y}\sum_{\hex}\left(\exp\left[-i\frac{2\pi\M}{\N}\left\{ \c_{\htwo}+\c_{\hthr}\right\} \right]\exp\left[i\frac{2\pi\L}{\N}\left\{ \r_{\htwo}+\r_{\hthr}\right\} \right]+H.c.\right)\,.\nonumber 
\end{eqnarray}
This model has a conserved quantity for each plaquette,
\begin{equation}
W_{\dfs}^{\,\,\hex}=\Y_{\hone}\ZZ_{\htwo}\X_{\hthr}\Y_{\hfou}\ZZ_{\hfiv}\X_{\hsix}=\exp\left(-i\frac{2\pi\M}{\N}\left[\c_{\hone}-\c_{\htwo}+\c_{\hfou}-\c_{\hfiv}\right]\right)\exp\left(i\frac{2\pi\L}{\N}\left[\r_{\hone}-\r_{\hthr}+\r_{\hfou}-\r_{\hsix}\right]\right)\,,
\end{equation}
where the $\M,\L$ factors are necessary for $W_{\dfs}^{\,\,\hex}$
to commute with the $J_{y}$-term. There are also conserved quantities
consisting of Weyl operators along any horizontal or 60-degree zig-zag
of the lattice,\begin{subequations}
\begin{eqnarray}
V_{\dfs}^{\leftrightsquigarrow} & = & \cdots\,\X_{\htwo}^{-1}\X_{\hthr}^{\phantom{-1}}\X_{\hfou}^{-1}\cdots=\exp\left(-\frac{2\pi}{\N}\left[\cdots\,\r_{\htwo}-\r_{\hthr}+\r_{\hfou}\cdots\right]\right)\\
V_{\dfs}^{\symx} & = & \cdots\,\ZZ_{\hone}^{\phantom{-1}}\ZZ_{\htwo}^{-1}\ZZ_{\hthr}^{\phantom{-1}}\cdots=\exp\left(\frac{2\pi}{\N}\left[\cdots\,\c_{\hone}-\c_{\htwo}+\c_{\hthr}\cdots\right]\right)\,.
\end{eqnarray}
\end{subequations}We have $V_{\dfs}^{\leftrightsquigarrow}V_{\dfs}^{\symx}=e^{-i\frac{4\pi}{N}}V_{\dfs}^{\symx}V_{\dfs}^{\leftrightsquigarrow}$,
implying a non-trivial ground-state degeneracy of the model on the
torus \cite{Barkeshli2015}. Unlike the toric code, this model is
not frustration-free and exhibits different phases for different values
of the parameters. We expand around the symmetric case $J_{x}\approx J_{y}\approx J_{z}$,
for which the $\N=2,\M=\L=1$ system is known to be gapless \cite{Kitaev2006}.

Performing the $\dfs\rightarrow\ft$ limit from Sec. \ref{subsec:Continuum-limit}
yields the coupled-oscillator Hamiltonian
\begin{equation}
H_{\ft}^{\text{hon}}\coloneqq\sum_{\hex}J_{x}\left(\x_{\hone}+\x_{\htwo}\right)^{2}+J_{y}\left(\p_{\htwo}-\x_{\htwo}+\p_{\hthr}-\x_{\hthr}\right)^{2}+J_{z}\left(\p_{\hthr}+\p_{\hfou}\right)^{2}\,.
\end{equation}
All conserved quantities are preserved: for each plaquette,
\begin{eqnarray}
W_{\ft}^{\,\,\hex} & = & \p_{\hone}-\x_{\hone}-\p_{\htwo}+\x_{\hthr}+\p_{\hfou}-\x_{\hfou}-\p_{\hfiv}+\x_{\hsix}\,,
\end{eqnarray}
and the string operators are
\begin{eqnarray}
V_{\ft}^{\symx} & = & \cdots\,\p_{\hone}-\p_{\htwo}+\p_{\hthr}\cdots\,\,\,\,\,\,\,\,\,\,\,\,\,\,\text{and}\,\,\,\,\,\,\,\,\,\,\,\,\,\,V_{\dfs}^{\leftrightsquigarrow}=\cdots\,\x_{\htwo}-\x_{\hthr}+\x_{\hfou}\cdots\,.
\end{eqnarray}
The string operators satisfy $[V_{\dfs}^{\leftrightsquigarrow},V_{\dfs}^{\symx}]=-2i$,
meaning that the ground-state degeneracy on a torus is infinite (since
there are no finite-dimensional irreducible representations of the
Heisenberg-Weyl algebra).

Performing the $\dfs\rightarrow\fs1$ limit from Sec. \ref{subsec:Rotor-limits}
yields the coupled-rotor system
\begin{eqnarray}
H_{\fs1} & = & -J_{x}\sum_{\hex}\cos\left(\L\left[\k_{\hone}+\k_{\htwo}\right]\right)-J_{z}\sum_{\hex}\cos\left(2\pi\Phi\left[\NN_{\hthr}+\NN_{\hfou}\right]\right)\\
 &  & -\frac{J_{y}}{2}\sum_{\hex}\left(\exp\left[-i2\pi\Phi\left\{ \NN_{\htwo}+\NN_{\hthr}\right\} \right]\exp\left[i\L\left\{ \k_{\htwo}+\k_{\hthr}\right\} \right]+H.c.\right)\,.\nonumber 
\end{eqnarray}
The conserved quantities turn into plaquette operators
\begin{eqnarray}
W_{\fs1} & = & \exp\left(-i2\pi\Phi\left[\NN_{\hone}-\NN_{\htwo}+\NN_{\hfou}-\NN_{\hfiv}\right]\right)\exp\left(i\L\left[\k_{\hone}-\k_{\hthr}+\k_{\hfou}-\k_{\hsix}\right]\right)
\end{eqnarray}
and string operator generators
\begin{eqnarray}
V_{\fs1}^{\symx} & = & \cdots\,\NN_{\hone}-\NN_{\htwo}+\NN_{\hthr}\cdots\,\,\,\,\,\,\,\,\,\,\,\,\,\,\,\text{and}\,\,\,\,\,\,\,\,\,\,\,\,\,\,\,V_{\fs1}^{\leftrightsquigarrow}=\cdots\,\k_{\htwo}-\k_{\hthr}+\k_{\hfou}\cdots\,.
\end{eqnarray}
The limit $\dfs\rightarrow\fs2$ is equivalent to this limit upon
$J_{x}\leftrightarrow J_{z}$ and a reflection.

\section{Summary and outlook\label{sec:d}}

The first part of this work provides an introduction to three types
of phase spaces whose conjugate variables can be related via a Fourier-type
transformation. The first type ($\dfs$) involves two discrete finite
conjugate variables $\{\c,\r\}$, the second type ($\fs$) involves
an integer-valued variable and its conjugate angle $\{\NN,\k\}$,
and the last type ($\ft$) is ordinary phase space of two continuous
conjugate variables $\{\x,\p\}$. The second part of this work is
concerned with converting Hamiltonians living in one phase space into
those in the others ($\dfs\rightarrow\ft$, $\dfs\rightarrow\fs$,
and $\fs\rightarrow\ft$). These limit-taking procedures correspond
directly to the limits connecting functions of a discrete periodic
variable ($\dfs$) to those of a continuous periodic variable ($\fs$)
or those of an unbounded variable ($\ft$) (Secs. 3.4.2 and 3.4.5
of Ref. \cite{kbwolf}, respectively). We outlined slightly generalized
versions of these well-known procedures and applied them to the $\dfs$
degrees of freedom in six models: the Harper equation, the Baxter
parafermionic spin chain, the Rabi model, the toric code, the Haah
cubic code, and the Kitaev honeycomb model. Interestingly, these straightforward
limit-taking procedures resulted in rotor and continuum limits for
all six models, making contact with the quantum Hall effect, optomechanics,
and lattice gauge theory. We hope these new models will be studied
further and that these techniques will be useful in generating interesting
new rotor and continuum limits. In particular, it would be interesting
to rigorously determine whether spectra of the $\dfs$ models converge
to their $\ft$ limits (as opposed to the $\ft$ limit being merely
a low-energy expansion). This has so far only been done for the oscillator,
but the solvability of most of the systems considered \cite{Baxter1989,Braak,Kitaev2003,Haah2011,Kitaev2006}
makes this direction quite promising.

We conclude with an outlook regarding experimental realizations. For
the Rabi model, implementation of the $\N\gg1$ circular structure
is admittedly tricky, as one has to engineer equal transition amplitudes
for all sites and also ensure that photons are either absorbed or
emitted, depending on the direction of the hopping. However, it is
not improbable that one can find a proper multi-level artificial atom
based on superconducting circuits whose transition rules satisfy the
$\N=3$ or even $\N=4$ Rabi models. The optimal candidate for realizing
high $\N$ cases however is optical lattices \cite{Amico2005a} or
trapped ions \cite{Noguchi2014,Tabakov2015,Li2017}, where ring-shaped
potentials can already be engineered. Similar experimental platforms
have been proposed to simulate $\Z_{\N}$ lattice gauge theories \cite{Notarnicola2015}.
For the many-body coupled rotor models, one could consider engineering
interactions between orbital angular momentum degrees of freedom of
photonic modes (e.g., \cite{Lvovsky2009,Nicolas2014}).
\begin{acknowledgments}
The authors thank Arpit Dua, William Sweeney, V. V. Sivak, Zlatko
K. Minev, and Liang Jiang for fruitful discussions and acknowledge
support from the Army Research Office. S.P. was partially supported
by INFN through the project ``QUANTUM''.
\end{acknowledgments}

\bibliographystyle{apsrev4-1t}
\bibliography{C:/Users/Victor/Dropbox/THESIS/library}

\end{document}